\documentclass[journal,compsoc, letterpaper]{IEEEtran}

\usepackage[cmex10]{amsmath}
\usepackage{cite,color,algorithm,algpseudocode,amsthm,amssymb,url,listings,subfigure,graphicx,setspace,url}

\usepackage[top=0.5in, bottom=.5in, left=.5in, right=.5in]{geometry}

\begin{document}
\title{The Rise of Social Botnets: Attacks and Countermeasures}

\author{Jinxue~Zhang,~\IEEEmembership{Student~Member,~IEEE,}
        Rui~Zhang,~\IEEEmembership{Member,~IEEE,}
        Yanchao~Zhang,~\IEEEmembership{Senior~Member,~IEEE,}
        and~Guanhua~Yan
\thanks{The preliminary version of this paper appeared in IEEE CNS 2013 \cite{ZhangOn13}.}%
\thanks{J. Zhang and Y. Zhang are with the Department of Electrical, Computer, and Energy Engineering, Arizona State University, Tempe, AZ 85287 (e-mail: {jxzhang, yczhang}@asu.edu).} %
\thanks{R. Zhang is with the Department of Electrical Engineering, University of Hawaii, Honolulu, HI 96822 (e-mail: ruizhang@hawaii.edu).} %
\thanks{G. Yan is with the Department of Computer Science, Binghamton University, New York, NY 13902 (e-mail: ghyan@binghamton.edu).}
}

\IEEEtitleabstractindextext{%

\begin{abstract}
Online social networks (OSNs) are increasingly threatened by \emph{social bot}s which are software-controlled OSN accounts that mimic human users with malicious intentions. A \emph{social botnet} refers to a group of social bots under the control of a single \emph{botmaster}, which collaborate to conduct malicious behavior while mimicking the interactions among normal OSN users to reduce their individual risk of being detected. We demonstrate the effectiveness and advantages of exploiting a social botnet for spam distribution and digital-influence manipulation through real experiments on Twitter and also trace-driven simulations. We also propose the corresponding countermeasures and evaluate their effectiveness. Our results can help understand the potentially detrimental effects of social botnets and help OSNs improve their bot(net) detection systems.
\end{abstract}


}

\maketitle

\IEEEraisesectionheading{\section{Introduction}}
\IEEEPARstart{O}{nline} social networks (OSNs) are increasingly threatened by \emph{social bot}s \cite{FerraRis14} which are software-controlled OSN accounts that mimic human users with malicious intentions. For example, according to a May 2012 article in Bloomberg Businessweek,\footnote{http://www.businessweek.com/articles/2012-05-24/likejacking-spammers-hit-social-media} as many as 40\% of the accounts on Facebook, Twitter, and other popular OSNs are spammer accounts (or social bots), and about 8\% of the messages sent via social networks are spams, approximately twice the volume of six months ago.  There have been reports on various attacks, abuses, and manipulations based on social bots \cite{FerraMan15}, such as infiltrating Facebook \cite{BoshmSoc11} or Twitter \cite{BilgeAll09, FreitRev14}, launching spam campaign \cite{GaoDet10,GrierSpa10, ThomaSus11}, and conducting political astroturf  \cite{RatkiDet11, RatkiTru11}.

A \emph{social botnet} refers to a group of social bots under the control of a single \emph{botmaster}, which collaborate to conduct malicious behavior while mimicking the interactions among normal OSN users to reduce their individual risk of being detected. For example, social bots on Twitter can follow others and retweet/answer others' tweets. Since a skewed following/followers (FF) ratio is a typical feature for social bots on Twitter \cite{WangDon10}, maintaining a balanced FF ratio in the social botnet makes it much easier for individual bots to escape detection. Creating a social botnet is also fairly easy due to the open APIs published by OSN providers. For example, we successfully created a network of 1,000 accounts on Twitter with \$57 to purchase 1,000 Twitter accounts instead of manually creating them.

Despite various studies \cite{GhoshSpa11, YangAna12,GhoshUnd12} confirming the existence of social botnets, neither have the greater danger from social botnets been unveiled nor have the countermeasures targeted on social botnets been proposed. In this paper, we first report two new social botnet attacks on Twitter, one of the most popular OSNs with over 302M monthly active users as of June 2015 and over 500M new tweets daily. Then we propose two defenses on the reported attacks, respectively. Our results help understand the potentially detrimental effects of social botnets and shed the light for Twitter and other OSNs to improve their bot(net) detection systems. More specifically, this paper makes the following contributions.

Firstly, we demonstrate the effectiveness and advantages of exploiting a social botnet for \emph{spam distribution} on Twitter. This attack is motivated by that Twitter currently only suspends the accounts that originate spam tweets without punishing those retweeting spam tweets \cite{TwitterRules}. If the social botnet is organized as a retweet tree in which only the root originates spam tweets and all the others merely retweet spams, all the social bots except the root bot can escape suspension. Given a set of social bots, we formulate the formation of the retweeting tree as a multi-objective optimization problem to minimize the time taken for a spam tweet to reach a maximum number of victim Twitter users at the lowest cost of the botmaster. Since the optimization is NP-hard, we give a heuristic solution and confirm its efficacy with real experiments on Twitter and trace-driven simulations.

Secondly, we show that a social botnet can easily manipulate the \emph{digital influence} \cite{ChaMea10,WengTwi10} of Twitter users, which has been increasingly used in ad targeting \cite{Chevy,KloutPerk12}, customer-service improvement \cite{KnappCS12}, recruitment \cite{KloutCEO}, and many other applications. This attack stems from the fact that almost all existing digital-influence tools such as Klout, Kred, and Retweet Rank, measure a user's digital influence exclusively based on his\footnote{No gender implication.} interactions with others users on Twitter. If social bots collaborate to manipulate the interactions of target Twitter users, they could effectively manipulate the victims' digital influence. The efficacy of this attack is confirmed by real experiments on Twitter.

Finally, we propose two countermeasures to defend against the two reported attacks, respectively. To defend against the botnet-based spam distribution, we maintain a spam score for each user and update the score whenever the corresponding user retweets a spam. The user is suspended if his spam score exceeds a predefined threshold. To defense against the botnet-based influence manipulation attacks, we propose to find sufficient credible users and only use the interactions originated from these credible users for digital-influence measurement. Moreover, based on the measurement in Section \S\ref{sec:diMani}, we design a new model to compute the influence score which is resilient to the manipulation from a single credible social bot. We confirm the effectiveness of both defenses via detailed simulation studies driven by real-world datasets.

The rest of the paper is organized as follows. \S\ref{sec:system} introduces the construction of a social botnet on Twitter. \S\ref{sec:spamCampaign} and \S\ref{sec:diMani} show the efficacy and merits of using the social botnet for spam distribution and digital-influence manipulation, respectively. \S\ref{sec:defend} details and evaluates two countermeasures. \S\ref{sec:relatedWork} discusses the related work. \S\ref{sec:conclusion} concludes this paper.

\section{Building a Social Botnet on Twitter}\label{sec:system}
%

In this paper, we focus on networked social bots in Twitter, so we first outline the Twitter basics to help illustrate our work. The readers familiar with Twitter can safely skip this paragraph without any loss of continuity. Unlike Facebook, the social relationships on Twitter are unidirectional by users \emph{following} others. If user $A$ follows user $B$, $A$ is $B$'s \textit{follower}, and $B$ is $A$'s \textit{friend}. In most cases, a user does not need prior consent from another user whom he wants to follow. Twitter also allows users to control who can follow them, but this feature is rarely used. In addition, users can choose to \emph{unfollow} others and \emph{block} their selected followers. A Twitter user can send text-based posts of up to 140 characters, known as \emph{tweet}s, which can be read by all its followers. Tweets can be public (the default setting) and are visible to anyone with or without a Twitter account, and they can also be protected and are only visible to previously approved Twitter followers. A \emph{retweet} is a re-posting of someone else's tweet. A user can retweet the tweets of anyone he follows or does not follow, and his retweets can be seen by all his followers. Moreover, a user can \emph{reply} to a post (tweet or retweet) and ensure that specific users can see his posts by \emph{mentioning} them via inserting ``@username'' for every specific user into his posts. Finally, each user has a \textit{timeline} which shows all the latest tweets, retweets, and replies of his followers.

We construct a social botnet on Twitter consisting of a botmaster and a number of social bots which are legitimate Twitter accounts. Twitter accounts can be manually created or purchased at affordable prices. For example, we bought 1,000 Twitter accounts with \$57 from some Internet sellers for experimental purposes only.
The botmaster is in the form of a Java application, which we developed from scratch based on  the OAuth protocol \cite{oauth} and open Twitter APIs. It could perform all the Twitter operations on behalf of all social bots to make the bots look like legitimate users. 

\section{Social Botnet for Spam Distribution}\label{sec:spamCampaign}

\subsection{Why the Social Botnet for Spam Distribution?}\label{sec:whyNSB}

As the popularity of Twitter rapidly grows, spammers have started to distribute spam tweets which can be broadly defined as unwanted tweets that contains malicious URLs in most cases or occasionally malicious texts \cite{GrierSpa10,ThomaSus11,Chudet12}. According to a study in 2010 \cite{GrierSpa10}, roughly 8\% of the URLs in tweets are malicious ones that direct users to scams/malware/phishing sites, and about 0.13\% of the spam URLs will be clicked. Given the massive scale of Twitter, understanding how spam tweets are distributed is important for designing effective spam defenses.

The simplest method for spam distribution is to let social bots distribute spam tweets independently from each other, which we refer to as the \emph{independent method}. In particular, the botmaster can instruct every bot to directly post spam tweets which can be seen by all its followers. According to the Twitter rules,\footnote{\url{http://support.twitter.com/articles/18311\#}} the accounts considered as spam originators will be permanently suspended. Since there are sophisticated techniques such as \cite{ThomaDes11,TwitterURL} detecting malicious URLs, this independent approach may subject almost all social bots to permanent suspension in a short time window.

A more advanced method, which we propose and refer to as the \emph{botnet method}, is to exploit the fact that Twitter currently only suspends the originators of spam tweets without punishing their retweeters. In the simplest case, the botmaster forms a single \emph{retweeting tree}, where every bot is associated with a unique vertex and is followed by its children bots. Then only the root bot originates spam tweets, and all the others simply retweet the spam tweets from their respective parent. Given the same set of social bots, both methods can distribute spam tweets to the same set of non-bot Twitter users, but only the root bot will be suspended under the botnet method. Obviously, the botnet method is economically beneficial for the botmaster because it involves non-trivial human effort or money to create a large social botnet.

We use an experiment on Twitter to validate our conjecture for the independent method.  Our experiment uses three different social botnets with each containing 100 bots. The experiment proceeds in hours. At the beginning of every hour, every bot in the same botnet almost simultaneously posts a spam tweet comprising two parts. The first part is different from every bot and randomly selected from the list of tweets returned after querying ``music,'' while the second part is an identical malicious URL randomly selected from the Shalla's blacklists (\url{http://www.shallalist.de/}) and shortened using the bitly service (\url{http://bitly.com}) for use on Twitter. We find that all the bots in the three botnets are suspected in two, five, and six hours. Based on this experiment, we can safely conjecture that the independent method will cause most bots in a larger botnet to be suspended in a short period, thus putting the botmaster at serious economic disadvantage.

We use a separate set of experiments to shed light on the advantage of the botnet method. In this experiment, we first use 111 bots to build a full 10-ary tree of depth two, i.e., each node except the leaves has exactly 10 children. The experiment proceeds in hourly rounds repeatedly on these 111 bots. At the beginning of every hour of the first round, the root bot posts a spam tweet, while all its descendants merely retweet the spam tweet after a small random delay. Then we replace the suspended bot by a random bot alive from the same network, re-organize the bot tree, and start the next round. We totally run the experiments for five rounds, in each of which only the root bot is suspended after six hours on average, and all other bots who just retweet the spams (with five times) remain alive. To check whether the bots will be suspended by retweeting more spams, we reduce the spamming frequency from one per hour to one per day, and repeat the experiment for ten more rounds, and all the retweeting bots were still alive at the end of the experiment. In addition, we use the similar methodology to test three other different botnets of 2, 40, and 100 bots, respectively, and obtain the similar results.

It has been very challenging in the research community to conduct experimental studies about the attacks on online social networks and also the corresponding countermeasures. In the experiments above, we have to control the social bots to post malicious URLs to evaluate Twitter's suspension policy, which may harm benign users. To minimize the negative impact on the legitimate users, we adopted a methodology similar to \cite{BoshmSoc11,ThomaSus11, MessiYou13, YangTas14}. Specifically, none of the purchased accounts followed any legitimate user and thus were very unlikely to be followed by legitimate users, which greatly reduced the possibility of the posted spams being viewed by legitimate users. In addition, we deleted every spam tweet immediately after the experiment to further avoid it being clicked by legitimate users. Our experiments clearly show that Twitter has a much more strict policy against posting original spam tweets than retweeting spam tweets.

\subsection{Optimal Social Botnet for Spam Distribution}\label{sec:modelNSB}

\S\ref{sec:whyNSB} motivates the benefits of using the social botnet for spam distribution on Twitter. Given a set of social bots, what is the optimal way for spam distribution? We give an affirmative answer to this important question in this section.

\subsubsection{Problem Setting and Performance Metrics}

We consider a botnet $\mathcal{V}$ of $n$ bots, where each bot $i\in [1,n]$ can be followed by other bots and also other Twitter users outside the botnet
(called non-bot followers hereafter). Let $\mathcal{F}_i$ denote the non-bot followers of bot $i$. Note that  $\mathcal{F}_i\cap
\mathcal{F}_j$ may be non-empty  ($\forall i\neq j$), meaning that any two bots may have overlapping non-bot followers. We further let
$\mathcal{F}=\bigcup_{i=1}^n{\mathcal{F}_i}$. How to attract non-bot followers for the bots is related to social engineering \cite{RealBoy}
and orthogonal to the focus of this paper. Note that it is very easy in practice for a bot to attract many non-bot followers, as shown in
\cite{BilgeAll09,BoshmSoc11,GhoshUnd12,YangAna12}.

The botmaster distributes spam tweets along one or multiple retweeting trees, and the vertices of every retweeting tree corresponds to a disjoint subset of the $n$ bots. In addition, every bot in a retweeting tree is followed by its children. As discussed, the root of every retweeting tree will originate spam tweets, which will appear in the Twitter timeline of its children bots and then be retweeted. The distribution of a particular spam tweet finishes until all the bots on all the retweeting trees either tweet or retweet it once and only once.

Given a set $\mathcal{V}$ with $n$ bots and $\mathcal{F}$, we propose three metrics to evaluate the efficacy of botnet-based spam distribution.

\begin{itemize}
    \item \emph{Coverage}: Let $\mathcal{C}$ denote the non-bot receivers of a given spam tweet and be called the \emph{coverage set}.
    The coverage of spam distribution is then defined as $\frac{|\mathcal{C}|}{|\mathcal{F}|}\in [0,1]$.

     \item \emph{Delay}: We define the delay of spam distribution, denoted by $\tau$, as the average time for each user in $\mathcal{C}$ to see a given spam tweet since it is generated by the root bot. A user may follow multiple bots and thus see the same spam tweet multiple times, in which case only the first time is counted.

     \item \emph{Cost}: We use $|\mathcal{S}|$ and $|\tilde{\mathcal{S}}|$ to define the cost of spam distribution, where $\mathcal{S}$ denotes the indices of suspended bots after distributing a given spam, and $\tilde{\mathcal{S}}$ denotes the set of non-bot followers will be lost due to the suspension of $\mathcal{S}$, i.e., $\tilde{\mathcal{S}}=\mathcal{C}\setminus (\bigcup_{i\in \mathcal{V\setminus S}}\mathcal{F}_i)$.
\end{itemize}

The above metrics motivates three design objectives. First, we obviously want to maximize the coverage to be one, which happens when all the $n$ bots participate in spam distribution by belonging to one retweeting tree. Second, many malicious URLs in spam tweets are hosted on compromised servers and will be invalidated once detected, and Twitter will remove spam tweets as soon as they are identified. It is thus also important to minimize the delay. Finally, since it incurs non-trivial human effort or money to create bots and attract followers for them, it is critical to minimize the cost as well.

\subsubsection{Design Constraints}\label{sec:designConstraint}

A major design challenge is how to circumvent Twitter's suspension rules\footnote{\url{http://support.twitter.com/articles/18311#}} that are evolving in accordance with changing user (mis)behavior. We classify the suspension rules into \textit{strict} and \textit{loose} ones. Violators of strict rules will be immediately suspended. The strict rule most relevant to our work is that the users originate spam tweets containing malicious URLs will be suspended. In contrast, a violator of loose rules will initially become suspicious and later be suspended if his violations of related loose rules exceed some unknown threshold Twitter defines and uses internally. Examples of loose rules include repeatedly posting others' tweets as your own or the same tweet, massively following/unfollowing people in a short time period, etc. In addition, the research community have discovered many useful loose rules for spam-tweet detection such as those in \cite{KwakWha10,LeeUnc10,StrinDet10,SongSpa11,ThomaDes11,YangDie11,ZhangDet11} which are likely to be or have been adopted by Twitter into their evolving suspension-rule list. As discussed, we use the botnet method for spam distribution in order to largely circumvent this strict rule. In the following, we introduce five design constraints related to some loose rules we consider most relevant. By following these constraints, the social bots can cause much less suspicion to Twitter and thus are much less likely to be suspended. 

1. The maximum height of a retweeting tree is $K= 10$ according to \cite{KwakWha10}. Hence we claim that any spam tweet will not be retweeted more than 10 times.

2.  A bot only retweets the spam tweets posted by its parent bot on the retweeting tree it follows, as retweeting the tweets from non-followed users is known to be effective in detecting spam tweets \cite{SongSpa11}.

3. Any spam tweet from an arbitrary bot will be retweeted by at most $100r$ percent of its followers. As $r$ approaches one, the bot will become increasingly suspicious according to community-based spam detection algorithms \cite{GaoTow12,Chudet12}. Recall that the followers of any bot $i\in [1,n]$ comprise other bots and also non-bot users (i.e., $\mathcal{F}_i$). Note that non-bot followers rarely retweet spam tweets in practice, but we require all bot followers to retweet spam tweets. Then bot $i$ can have no more than $\lceil\frac{r|\mathcal{F}_i|}{1-r}\rceil$ bot followers.

4. The retweeting lag at any hop $j\in [1,K]$ is a random variable $t_i$ which follows a hop-wise statistical distribution according to \cite{KwakWha10}, as it is quite abnormal for a user to immediately retweet a post once seeing it. Here the retweeting lag is defined as the time elapsed when a bot sees a spam tweet until it retweets it.

5. The social bots within the first $M$ hops will be suspended once Twitter finds that they are involved in (re)tweeting a spam tweet. This constraint is motivated by recent findings \cite{YangAna12} that spammer accounts on Twitter tend to be connected and clustered by mutual followings. It is thus reasonable to assume that Twitter either have been utilized or will soon utilize these research findings to suspend the accounts involved in distributing a spam tweet within the first $M>0$ hops. After introducing this constraint, we relax the third one by allowing arbitrary topology in the first $M$ hops because all of its bots will be suspended.

\subsubsection{Problem Formulation}

\begin{figure}[t]
\centering
\includegraphics[width=0.35 \textwidth]{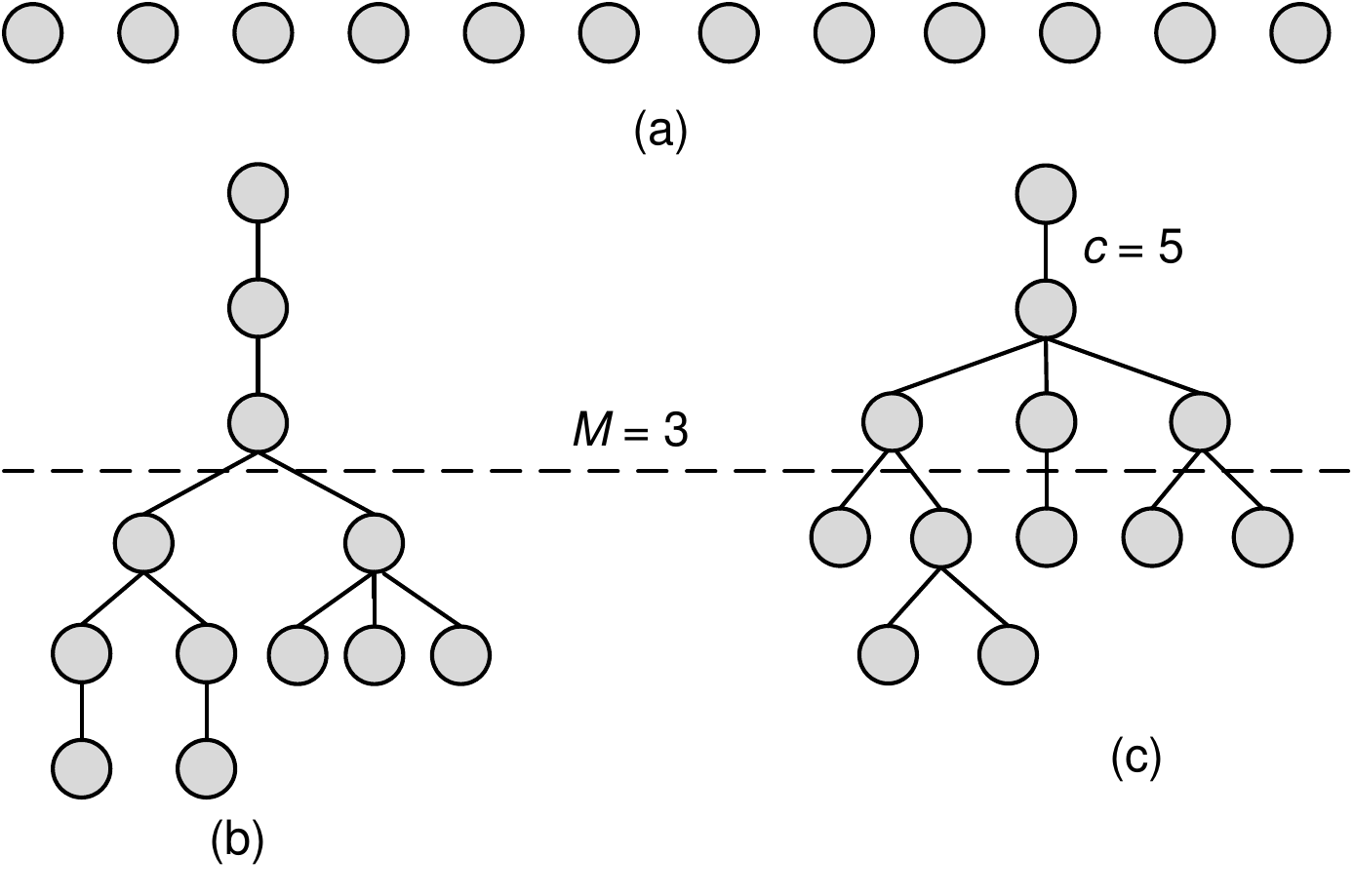}
\caption{Exemplary retweeting trees with 12 bots, where $M=3$ and the botmaster's suspension budget is $c=5$.}
\label{fig:retweet-forest}
\end{figure}

Give the above design objectives and constraints, we now attempt to formulate botnet-based spam distribution as an optimization problem. The major challenge lies in the infeasibility of simultaneously achieving the maximum coverage, the minimum delay, and the minimum cost. Fig. ~\ref{fig:retweet-forest} shows an example with 12 bots and $M=3$, and we assume that every bot has the same number of non-bot followers. In one extreme shown in Fig.~\ref{fig:retweet-forest}(a), we can minimize the delay $\tau$ by letting every bot be a root bot, but the cost is obviously the maximum possible because all the bots will be suspended. In the other extreme shown in Fig.~\ref{fig:retweet-forest}(b), we can form a single retweeting tree with exactly three bots within the first three hops, in which case we can achieve the minimum possible cost, but the achievable delay will always be larger than that in the first case no matter how the retweeting tree beyond three hops is formed.  In addition, we assume for the second case that the retweeting tree can include all the 12 bots, leading to the same coverage of one as in the first case. If there are too many bots, however, some of them may not be able to be incorporated into the retweeting tree due to  the first and third design constraints, and the resulting coverage will be smaller than that of the first case.

To deal with the above challenge, assume that the botmaster has a suspension budget $c \in [M,n]$ bots, referring to the maximum number of suspended bots it can tolerate. Note that the more bots in the first $M$ hops, the more non-bot followers in $\mathcal{F}$ closer to the root bot which can receive a given spam tweet in shorter time, and thus the smaller the delay. Under the budget constraint, the minimum delay can hence be achieved only when there are exactly $c$ bots within the first $M$ hops, as shown in Fig.~\ref{fig:retweet-forest}(c) with $c=5$.

What is the optimal way to form a retweeting tree as in Fig.~\ref{fig:retweet-forest}(c) given the cost, coverage, and delay requirements? Recall that the cost is defined by $|\mathcal{S}|$ and $|\tilde{\mathcal{S}}|$. Since $|\mathcal{S}|=c$ under the budget constraint, we just need to minimize $|\tilde{\mathcal{S}}|$. To mathematically express the cost and coverage requirements, we let $\{\mathcal{V}_k\}_{k=1}^K$ denote $K$ disjoint subsets of the bot indices $\{1,\dots,n\}$, where $K=10$ is the maximum height of the retweeting tree (see Constraint~1), $\mathcal{V}_k$ denote the bot indices at level $k$ of the retweeting tree, and $\bigcup_{k=1}^K \mathcal{V}_i \subseteq \{1,\dots,n\}$. If the optimal retweeting tree eventually found is of depth $K^\ast<K$, the sets $\{\mathcal{V}_k\}_{k=K^\ast+1}^K$ will all be empty. Recall that $\mathcal{F}_i$ denotes the set of non-bot followers of bot $i\in [1,n]$ and that $\mathcal{F}=\bigcup_{i=1}^n{\mathcal{F}_i}$. Then we have $\tilde{\mathcal{S}}=\mathcal{C} \setminus (\bigcup_{i\in \mathcal{V}_k, k\in[M+1,K^\ast]}\mathcal{F}_i) $ and the coverage set $\mathcal{C}=\bigcup_{i\in \mathcal{V}_k, k\in[1,K]}\mathcal{F}_i\subseteq \mathcal{F}$ and need to maximize $|\mathcal{C}|$. Since $\tilde{\mathcal{S}}\subseteq \mathcal{C}$, we can combine the cost and coverage requirements into a single metric $\frac{|\tilde{\mathcal{S}}|}{|\mathcal{C}|}$ and then attempt to minimize it.

It is a little more complicated to derive the delay. As discussed, a non-bot user may follow multiple bots at different levels, in which case it is considered a follower on the lowest level among those. Let $\Phi_k$ denote the set of non-bot followers at $k\in [1,K]$. It follows that $\Phi_1=\mathcal{F}_i$ ($i\in \mathcal{V}_1$) and $\Phi_k=\bigcup_{i\in \mathcal{V}_k}{\mathcal{F}_i}-\bigcup_{l=1}^{k-1}{\Phi_l}$ for $k\in [1,K]$.
According to Constraint 4, it takes $\sum_{j=1}^{k}t_j$ for a spam tweet to reach level-$k$ non-bot followers, where $t_j$ denotes the retweeting lag of hop $j\in[1,K]$, and $t_1=0$. Since there are totally $|\Phi_k|$ non-bot followers at level $k$ and $|\mathcal{C}|$ non-bot followers across all levels, we can compute the delay as
\[\tau=\frac{1}{|\mathcal{C}|} \sum_{k=1}^K \sum_{j=1}^{k}t_j |\Phi_k|\;.\]

Finally, we reduce the three-objective optimization problem to the following single-objective minimization problem.
\begin{equation}\label{opt:spam}
\begin{split}
\min &\quad f(\{\mathcal{V}_k\}_{k=1}^K)=\alpha \beta\frac{|\tilde{\mathcal{S}}|}{|\mathcal{C}|}+ (1-\alpha) \tau \\
\textrm{s.t.} &\quad \bigcup_{k=1}^K \mathcal{V}_i \subseteq \{1,\dots,n\}\\
&\quad \mathcal{V}_i\cap \mathcal{V}_j=\phi, \forall i\neq j\in [1,K]\\
&\quad |\bigcup_{k=1}^M \mathcal{V}_i| \leq c\\
&\quad \sum_{i\in\mathcal{V}_k} \lceil\frac{r|\mathcal{F}_i|}{1-r}\rceil \geq |\mathcal{V}_{k+1}|, \forall k\in[M-1,K-1]\\
\end{split}
\end{equation}
We have two remarks here. First, $\alpha \in [0,1]$ is a adjustable weight that reflects the relative importance of coverage and delay, and $\beta$ is a fixed scaling factor to unify two different objective units. Second, the last constraint is due to the aforementioned third design constraint.

The above optimization problem can be viewed as a variation of classical set partition problem (SPP), which is NP-hard. In what follows, we introduce a heuristic approximation solution by constructing a collection of disjointed subsets $\{\mathcal{V}_k\}_{k=0}^K$ from the botnet set.

\begin{figure*}
\centering
 \subfigure[$\alpha=0.4$]{\label{fig:sp-alpha0.4}\includegraphics[width=0.28\textwidth]{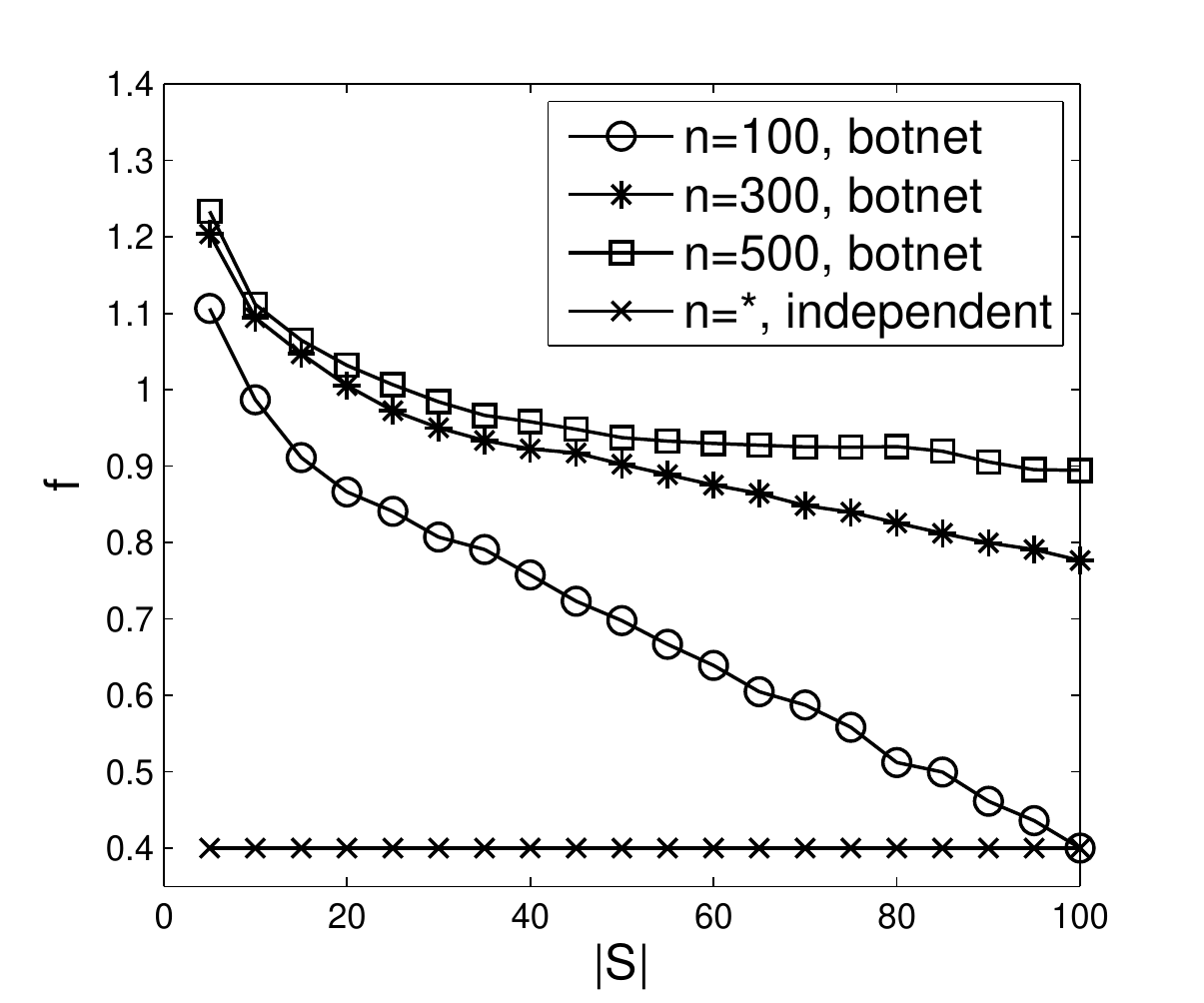}}\hfill
    \subfigure[$\alpha=0.65$]{\label{fig:sp-alpha0.65}\includegraphics[width=0.28\textwidth]{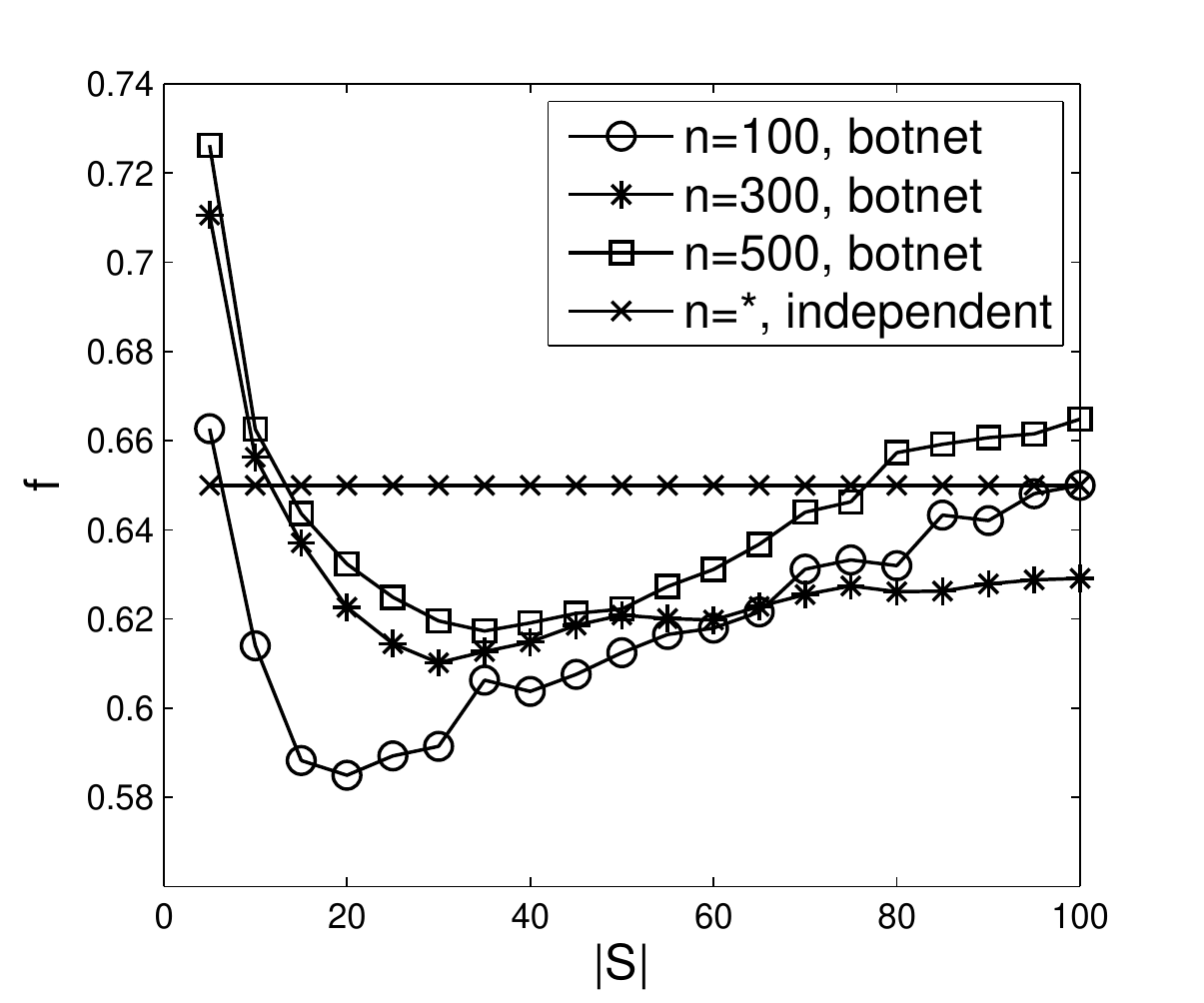}}\hfill
    \subfigure[$\alpha=0.9$]{\label{fig:sp-alpha0.9}\includegraphics[width=0.28\textwidth]{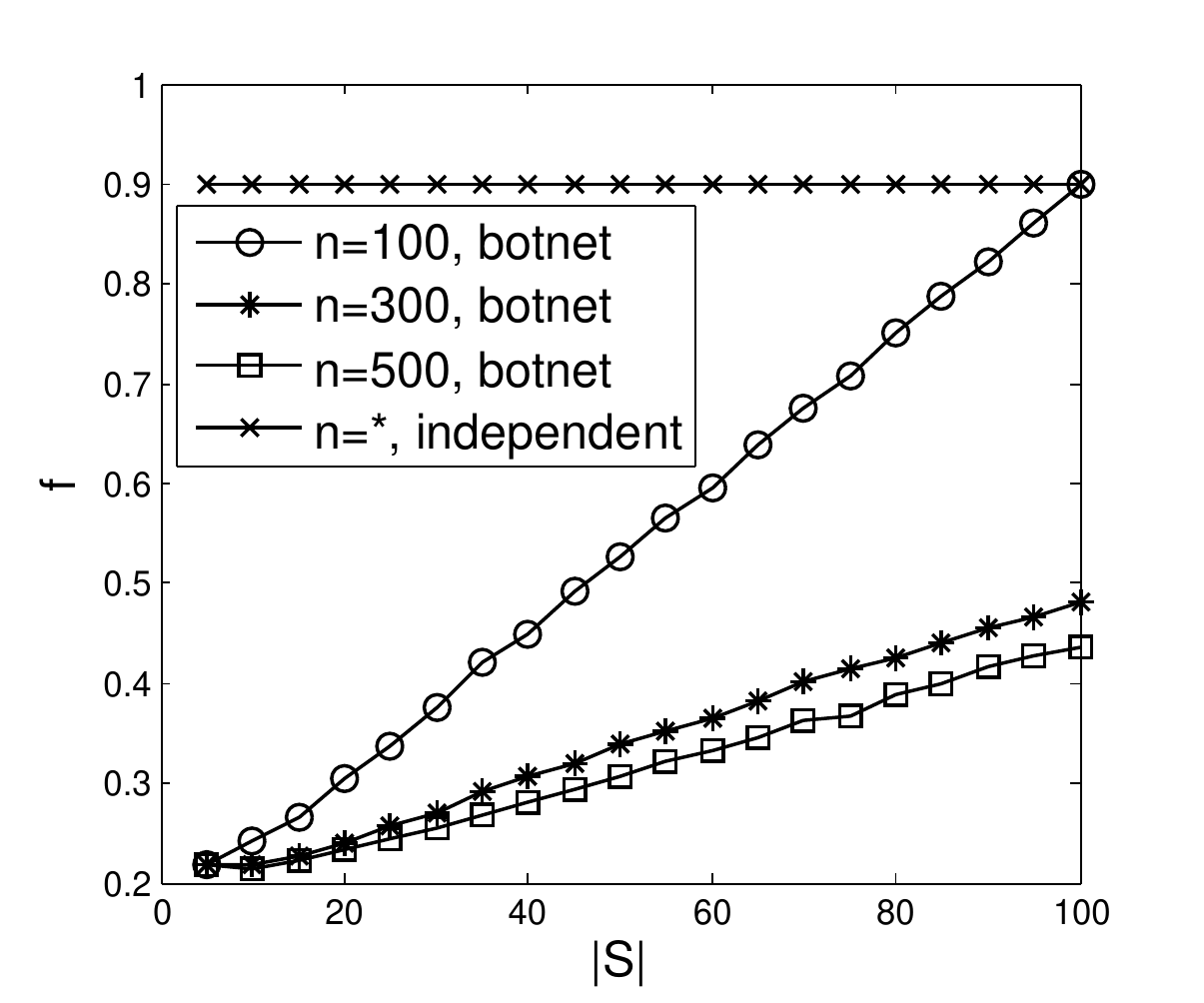}} \hfill
\caption{Performance comparison of independent and botnet methods in spam distribution at different $\alpha$s in terms of the single objective $f$.}
\label{fig:sp-AOF}
\end{figure*}

\begin{figure}
\centering
 \subfigure[\#lost legitimate followers]{\label{fig:sp-S2}\includegraphics[width=0.24\textwidth]{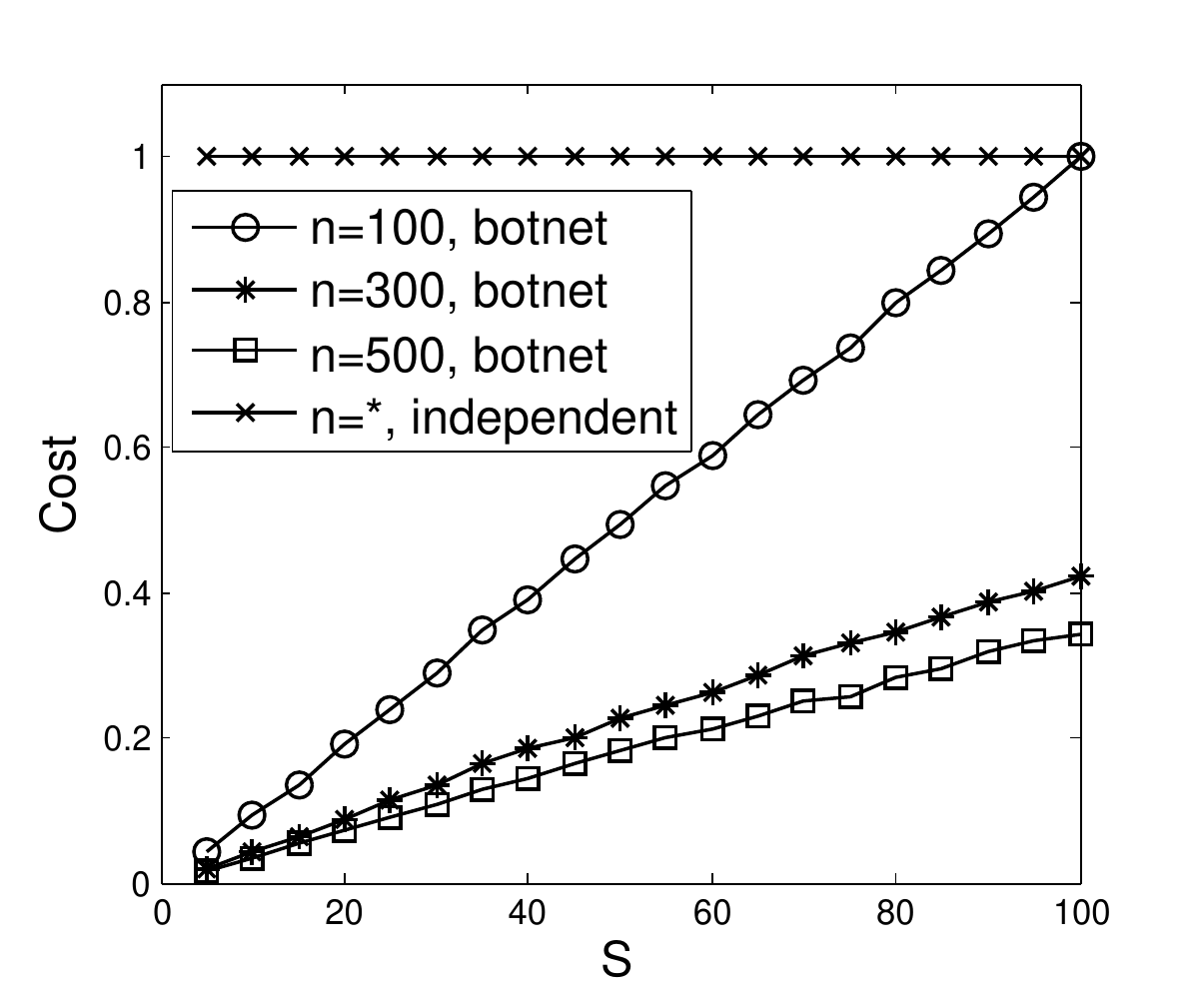}}\hfill
   \subfigure[Average delay $\tau$ (hours)]{\label{fig:sp-T}\includegraphics[width=0.24\textwidth]{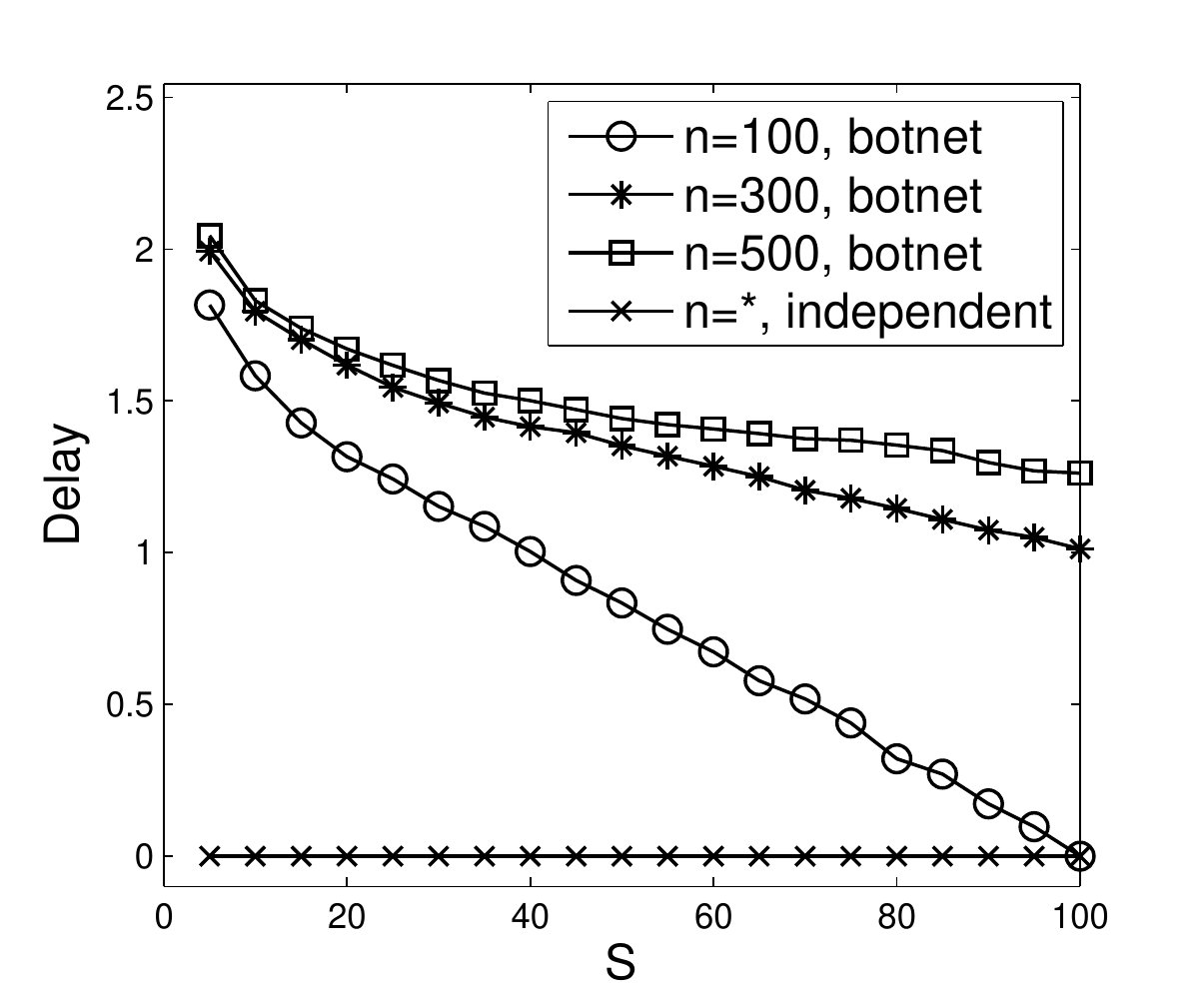}}
    \caption{Performance comparison of independent and botnet methods in spam distribution in terms of separate objective.}
    \label{fig:sp-CST}
\end{figure}

\subsubsection{Heuristic Solution}\label{sec:heusolution}
Our intuition is to use all the budget $c$ and fill the first $M$ hops of the retweeting trees with the bots having the lowest suspension
cost in terms of the number of lost non-bot followers. We then recursively place the bots with the highest number of non-bot followers from
level $M+1$ to the last level, in order to reduce the average latency as well as cost.

To begin with, our approximation is built on the solutions to the traditional maximum coverage problem (or \textsc{MaxCover}) and the minimum
coverage problem (\textsc{MinCover}), which is to select $k$ sets from a collection of subsets of a ground set so that the their union is maximized \cite{ChierMax10} or minimized, respectively. \textsc{MaxCover} and \textsc{MinCover} problems are both NP-hard and have greedy approximation algorithms by iteratively selecting the maximum or minimum subset after extracting the selected elements, respectively.

Our solution consists of the following two steps. First, given the budget $c$ for $\mathcal{S}$ and that all the followers in $\tilde{\mathcal{S}}$ will be lost because of the suspension, we minimize the objective $|\tilde{\mathcal{S}}|$ by using \textsc{MinCover}
to choose $c$ bots as $\tilde{\mathcal{S}}$ with the minimum total number of non-bot followers. In doing so, we can determine the union of the bot set for the first $M$ level. The bot subset in each level will be determined later. Here we just assume that $\tilde{\mathcal{S}}$ has been divided into the $M$ groups, each corresponding to the bots in one of the first $M$ levels. Second, we construct $\{\mathcal{V}_k\}_{k=M+1}^K$ to greedily increase the coverage $C$ and at the same time lower the average delay $T$. Specifically, assuming that we have known $\mathcal{V}_M$, to determine $\mathcal{V}_{M+1}$, we first set the cardinality of $\mathcal{V}_{M+1}$ be equal to $\sum_{i\in\mathcal{V}_k} \lceil\frac{r|\mathcal{F}_i|}{1-r}\rceil $ according to the last constraint in (\ref{opt:spam}) and then use \textsc{MaxCover} to choose a subset of $|\mathcal{V}_{M+1}|$ bots from the remaining bot set with the maximum number of non-bot followers. We repeat this greedy selection for every level $k=M+2, \ldots, K$.

The remaining problem is how to partition $\tilde{\mathcal{S}}$ into $M$ subsets, each corresponding to one of the first $M$ levels. A heuristic observation here is that we need to maximize $|\mathcal{V}_M|$, as the more non-bot followers of the social bots in the $M$th level, the more social bots in level $M+1$ and subsequent levels, and also the lower average delay according to the last constraint in
(\ref{opt:spam}). Given the budget $|\tilde{\mathcal{S}}|=c$, we obtain $|\mathcal{V}_M|_{max}=c-M+1$ when the retweeting forest has a single tree whose first $M-1$ levels form a straight line, as shown in Fig.~\ref{fig:retweet-forest}(b). The bots in the $M$th level is then determined by using \textsc{MaxCover} to choose the $c-M+1$ bots from $\tilde{\mathcal{S}}$ with the maximum number of non-bot followers.

To determine the level for each of the remaining $M-1$ bots, we sort the remaining $M-1$ bots in $\tilde{\mathcal{S}}$ in the descending
order according to the number of their non-bot followers and assign them to the corresponding level, e.g., the bot with the highest number
of non-bot followers will be assigned to the first level. Note that it is possible that after we maximizing the number of bots at the
$M$th level, the remaining bots are less than the allowed on the $M$th level, so the ($M+1$)-th is not full. To further reduce the average delay in such cases, we move the exceeding bots in the $M$th level to the first level.

After determining $\{\mathcal{V}_i\}_{i=1}^K$ from the social-bot set $\{1,\dots,n\}$, we can then build the final retweeting forest
(tree). Specifically, the number of retweeting trees is equal to the cardinality of $\mathcal{V}_1$, which is one if the ($M+1$)-th level
is full or greater than one otherwise. We then randomly choose one social bot from  $\mathcal{V}_1$ to be the root of the retweeting tree
with more than one level, which is followed by the bots from the second to $M$th level determined by $\mathcal{V}_2$ to $\mathcal{V}_M$,
respectively. Finally, we build the level from $k=M+1$ to $K$ by selecting certain number of social bots from $\mathcal{V}_k$ according the
last constraint in Eq.~(\ref{opt:spam}).

\subsection{Trace-driven Evaluation}\label{sec:evaluationNSB}

We conduct trace-driven simulations to compare the performance of spam distribution using the independent and botnet methods, as well as evaluating the tradeoffs among the multiple goals in the botnet method.

The evaluation for independent bots is straightforward. In particular, given the bot set $\mathcal{V}$ with $|\mathcal{V}|=n$, we place all the bots in the first level which will be suspended completely. We then have $\mathcal{C}=\tilde{\mathcal{S}}=n$, and $\tau=0$. The single objective in Problem (\ref{opt:spam}) is thus $f=\alpha$.

To evaluate the botnet method, we set up the simulations according to existing measurement data and heuristics. We set $K = 10$, and $t_1=0,t_i = 0.5i$ hour for $i=2,\ldots K$ according to \cite{KwakWha10}. To build $\{\mathcal{F}_i\}_{i=1}^n$, we generate $|\mathcal{F}_i|$ according to the Gaussian distribution with $\mu=32$ as the average number of followers in the dataset of \cite{KwakWha10}. We also set the variance to $\sigma^2=5$, generate a legitimate follower set $\mathcal{F}$ with $|\mathcal{F}| = 6000$,\footnote{For the Gaussian distribution with $\mu=32$ and $\sigma^2=5$, the probability for generating a negative $|\mathcal{F}_i|$ is negligible.} and randomly choose $|\mathcal{F}_i|$ followers from the set $\mathcal{F}$ for each bot $i$. In addition, according to \cite{YangAna12}, the average path length of the spammer community is 2.60, so we set $M=3$ to suspend the bots in the first three hops of $\mathcal{F}$. Finally, we set $\beta$ to one and the retweeting ratio $r=0.2$. Due to space constraints, we focus on on the impact of $\alpha$, $c,$ and $n$ and do not report the impact of $|\mathcal{F}|$, $\sigma^2$, $r$, $M,$ or $\beta$ in this paper.

Fig.~\ref{fig:sp-AOF} compares the independent and botnet methods using the objective function $f$  with different weights $\alpha$. As stated before, $f$ is simply equal to $\alpha$ for the independent case because $\tau=0$ and $\tilde{\mathcal{S}}=\mathcal{C}$. For the botnet method, the  objective $f$ is the weighted sum of $\frac{|\tilde{\mathcal{S}}|}{|\mathcal{C}|}$ and the delay $\tau$. When $\alpha$ is small, $\tau$ has higher impact on $f$ than $\frac{|\tilde{\mathcal{S}}|}{|\mathcal{C}|}$, while when $\alpha$ is large, $\frac{|\tilde{\mathcal{S}}|}{|\mathcal{C}|}$ will dominate $f$. Specifically, we can see from Fig.~\ref{fig:sp-alpha0.4} that when $\alpha = 0.4$, the independent method outperforms the botnet method with smaller $f$. However, as shown in Fig.~\ref{fig:sp-alpha0.65}, when $\alpha$ increases to 0.65,  the botnet method can achieve lower $f$ than the independent method does. This trend is more obvious for the same reason in Fig.~\ref{fig:sp-alpha0.9} where $\alpha = 0.9$. 


Figs.~\ref{fig:sp-CST} compare the two methods in terms of separate objectives including the number of lost legitimate followers and the average delay under different budget $|\mathcal{S}|=c$. We can see that both methods have the same coverage $|\mathcal{C}|$, which is equal to $|\mathcal{F}|$, as well as the maximum value of $C$. In addition, we can see from Fig.~\ref{fig:sp-T} that the delay of the independent method is zero, while that of botnet method could be on the order of hours. Finally, Fig.~\ref{fig:sp-S2} shows that the botnet method has significant advantage  than the independent method in terms  of  $|\tilde{\mathcal{S}}|$, the number of lost legitimate followers, as $|\tilde{\mathcal{S}}|$ is always equal to $|\mathcal{F}|$ for the independent scheme.

\begin{table*}
\centering
\footnotesize
\begin{tabular}{l*{6}{c}p{5.5cm}}
&&&&\multicolumn{3}{c}{Digital-influence score}\\
\cline{5-7}
Vendor              & Start & \#users (M) &  Update & Scale & Average & 90-percentile & Target OSNs\\
\hline
Klout       & 2008 & 620+ & daily & 0-100 & 20 & 50 & Twitter, Facebook, LinkedIn, Google+, Instagram, Foursquare\\
Kred        & 2011 & - & hourly & 0-1000 & 500 & 656 & Twitter, LinkedIn, Facebook, Quora \\
PeerIndex   & 2009 & 50+    & daily     & 0-100 & 19 & 42 & Twitter, Facebook\\
Retweet Rank        & 2008 & 3.5& hourly & 0-100 & 50 & 90  & Twitter\\
Tweet Grader        & 2010 & 10+ & daily & 0-100 & 50 & 90 & Twitter\\
Twitalyzer      & 2009 & 1 & daily & 0-100 & 50 & 90 & Twitter \\
\end{tabular}
\caption{Six popular digital-influence software vendors. The data was collected at October, 2014.}
\label{tbl:vendors}
\end{table*}


Finally, the botnet size $n$ also has some impact on separate objectives in the botnet case. Fig.~\ref{fig:sp-S2} shows that $\frac{|\tilde{\mathcal{S}}|}{|\mathcal{C}|}$ decreases as $n$ increases. The reason is that the larger $n$, the more bots with less non-bot followers will be assigned to the first $M$ levels, resulting in smaller $|\tilde{\mathcal{S}}|$ and thus larger $\frac{|\tilde{\mathcal{S}}|}{|\mathcal{C}|}$.  In addition, Fig.~\ref{fig:sp-T} shows that the larger $n$, the higher the average delay $\tau$, which is also expected.

In summary, from the view point of the botmaster, these evaluations show that the botnet scheme is more flexible than the independent method when considering multiple objectives of the spam distribution at the same time.

\section{Social Botnet for Digital-influence Manipulation}\label{sec:diMani}
In this section, we first briefly introduce digital influence and then experimentally show the efficacy of using the social botnet to manipulate digital influence.

\subsection{Rise of Digital Influence}\label{sec:DI}

Digital influence is one of the hottest trends in social media  and is defined as ``the ability to cause effect, change behavior, and drive measurable outcomes online'' in \cite{SolisRis12}.  The huge commercial potential of digital influence is in line with the increasingly recognized importance of word-of-mouth marketing on social networks. There are also growing business cases in which various companies successfully promoted their services/products by reaching out to most influential social-network users in their respective context \cite{SolisRis12}.

The future of digital influence also relies on effective tools to measure it. As reported in \cite{SolisRis12}, there are over 20 popular digital-influence software vendors such as Klout \cite{Klout} , Kred \cite{Kred}, Retweet Rank \cite{retweetrank}, PeerIndex \cite{peerindex}, TwitterGrade \cite{tweetgrader} and Twitalyzer \cite{twitalyzer}. Every vendor has its proprietary method to compute an \emph{influence score} for a given user based on his activities within his affiliated social network such as Twitter, Facebook, Google+, and LinkedIn, and higher scores represent greater influence. As shown in Table \ref{tbl:vendors}, Klout, Kred, and PeerIndex use normalized scores with different average values and scales, while RetweetRank, TweetGrader, and Twitalyzer represent digital-influence scores using percentile.

The typical business model of digital-influence vendors is based around connecting businesses with individuals of high influence. Companies have paid to contact individuals with high influence scores in hopes that free merchandise and other perks will influence them to spread positive publicity for them. For example, in 2011 Chevy offered 139 3-day test drives of its 2012 Sonic model to selected participants with the Klout score of at least 35 \cite{Chevy}. As another example, it has been reported that some recruiters have used the digital-influence scores to select qualified candidates \cite{KloutCEO}. In addition, customer service providers like Genesys  prioritize customer complaints according to their digital-influence scores to avoid the amplification of complaints by influential users in OSNs \cite{KnappCS12}. Klout announced a growth of 2,000 new partners over a one year period in May 2012.

\subsection{Botnet-based Digital-influence Manipulation}\label{sec:BotDIM}

Given the great potential of digital influence, whether it can be maliciously manipulated is an important research issue.  For example, assume that malicious users could collude to significantly increase their influence scores. A company using the digital-influence service may consider them most influential and choose them as the targets of important marketing campaigns by mistake, thus having potentially huge financial loss, while malicious users can potentially benefit, e.g., by getting free sample products. In addition, malicious users may attract more legitimate followers who tend to follow most influential users and thus become more influential.

As the first work of its kind, we now explore the feasibility of using the botnet to manipulate digital influence. Our studies involve three most popular digital-influence vendors for Twitter users: Klout, Kred, and Retweet Rank. For clarity, we summarize their key features as follows.

\begin{itemize}
    \item Klout: The Klout score of a Twitter user is on the scale of 1 to 100 and updated daily based on how frequently he or she is retweeted and mentioned in the last 90 days. The average Klout score is close to 20, and the score of the 90th percentile is over 50.\footnote{\url{http://therealtimereport.com/2012/04/11/how-good-is-your-klout-score/}}
     \item Kred: The Kred score of a Twitter user is on the scale of 1 to 1,000 and updated in real time according to how frequently he or she is retweeted, replied, mentioned, and followed on Twitter in the last 1,000 days.\footnote{\url{http://kred.com/rules}}
     \item Retweet Rank: It ranks the users based on how many times they each have been retweeted recently and how many followers/friends they each have.\footnote{http://www.retweetrank.com/view/about} Retweet ranks are updated on an hourly basis, and a retweet rank of $x$ means that the corresponding user is the $x$th most influential on Twitter. A retweet rank can also be translated into a percentile score ranging from 1 to 100, indicating how the user score relative to other Twitter users.
\end{itemize}

Given a social botnet of $n$ bots, we want to investigate whether it is feasible to generate an arbitrary influence score $d_i$ for every bot $i\in[1,n]$ under each of the above three tools. Since every bot is usually indistinguishable from a legitimate user, our investigation can also shed light on the feasibility of using the botnet to manipulate the influence score of an arbitrary Twitter user. Since every digital-influence vendor (including the above three) usually keeps confidential its detailed algorithm for computing influence scores, our studies are purely based on real Twitter experiments. According to \cite{SolisRis12}, we conjecture  that the following three factors play the important role in determining a user's digital-influence score.

\begin{itemize}
     \item \emph{Actions}. Both the number and the type of actions have large impacts on a user's digital-influence score. Intuitively, the more actions the user can attract, the higher his digital-influence score. Moreover, different types of actions may have different impacts. For example, retweeting or replying should be more indicative than following because the latter has been shown to be more vulnerable to fake \cite{YangAna12}.
     \item \emph{Audiences}. Given a target user $u$, we define all the users who have retweeted or replied $u$'s tweets, or mentioned or followed $u$ as $u$'s \emph{audiences}. We conjecture that the larger the audience size, the higher the digital-influence scores. The intuition is that the target user is more influential if each of his 100 tweets is retweeted by a different user than all 100 tweets are retweeted by the same one user. We also conjecture that the higher the digital-influence scores his audience have, the higher the target user's digital-influence score.
     \item \emph{Popularity of tweets}. The digital-influence score of a target user is determined by the popularity of his tweets. We want to explore how the distribution of the tweet popularity determines the overall influence of the target user.
\end{itemize}

Based on these three factors, we then present how to orchestrate the social botnets to manipulate the digital-influence scores.

%

\begin{figure*}
\centering
\begin{minipage}{0.48\textwidth}
\centering
\subfigure[Klout Scores]{\label{fig:di-audi-klout}\includegraphics[width=0.5\textwidth]{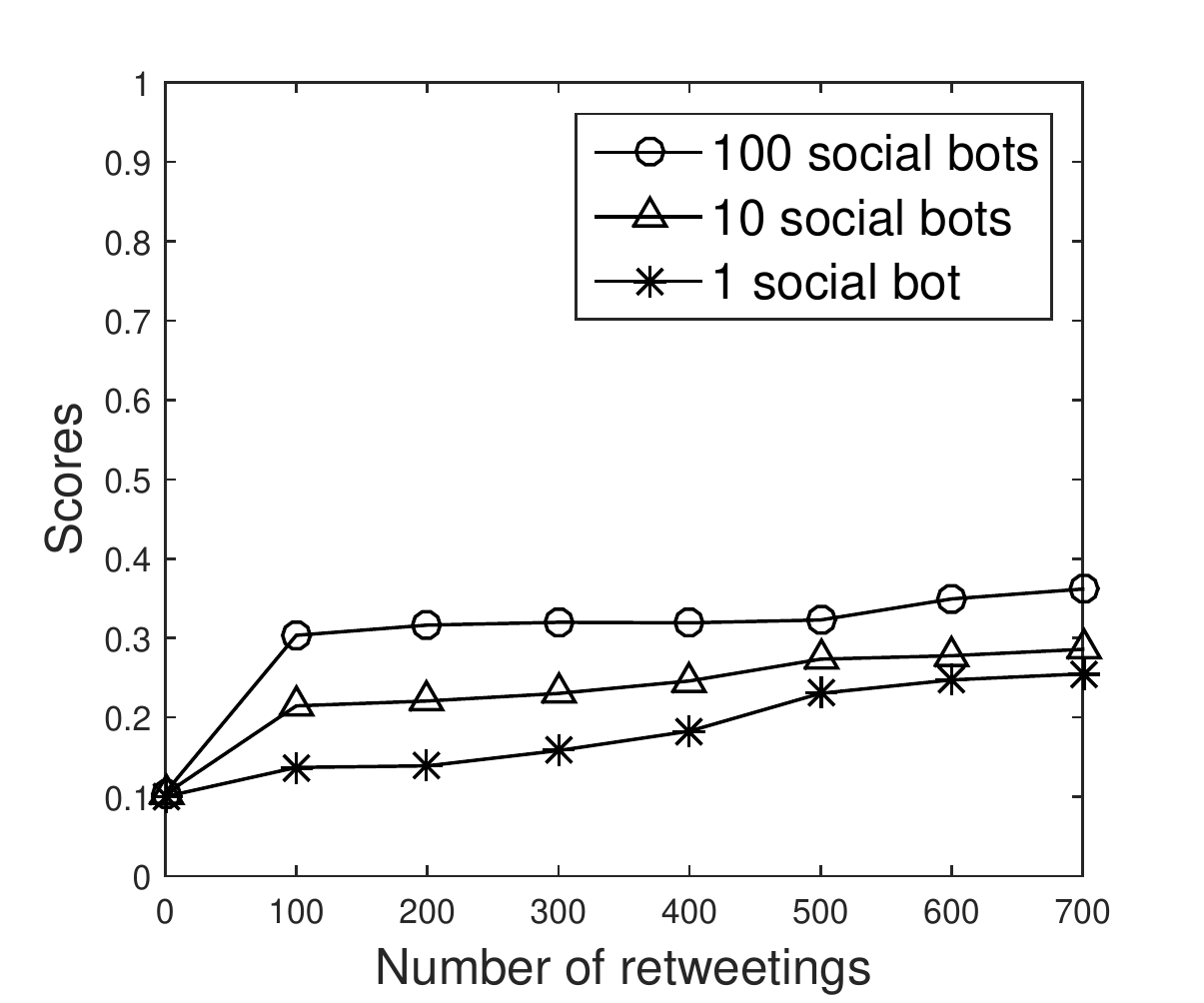}}\hfill
\subfigure[Kred and Retweet Rank Scores]{\label{fig:di-audi-kred}\includegraphics[width=0.5\textwidth]{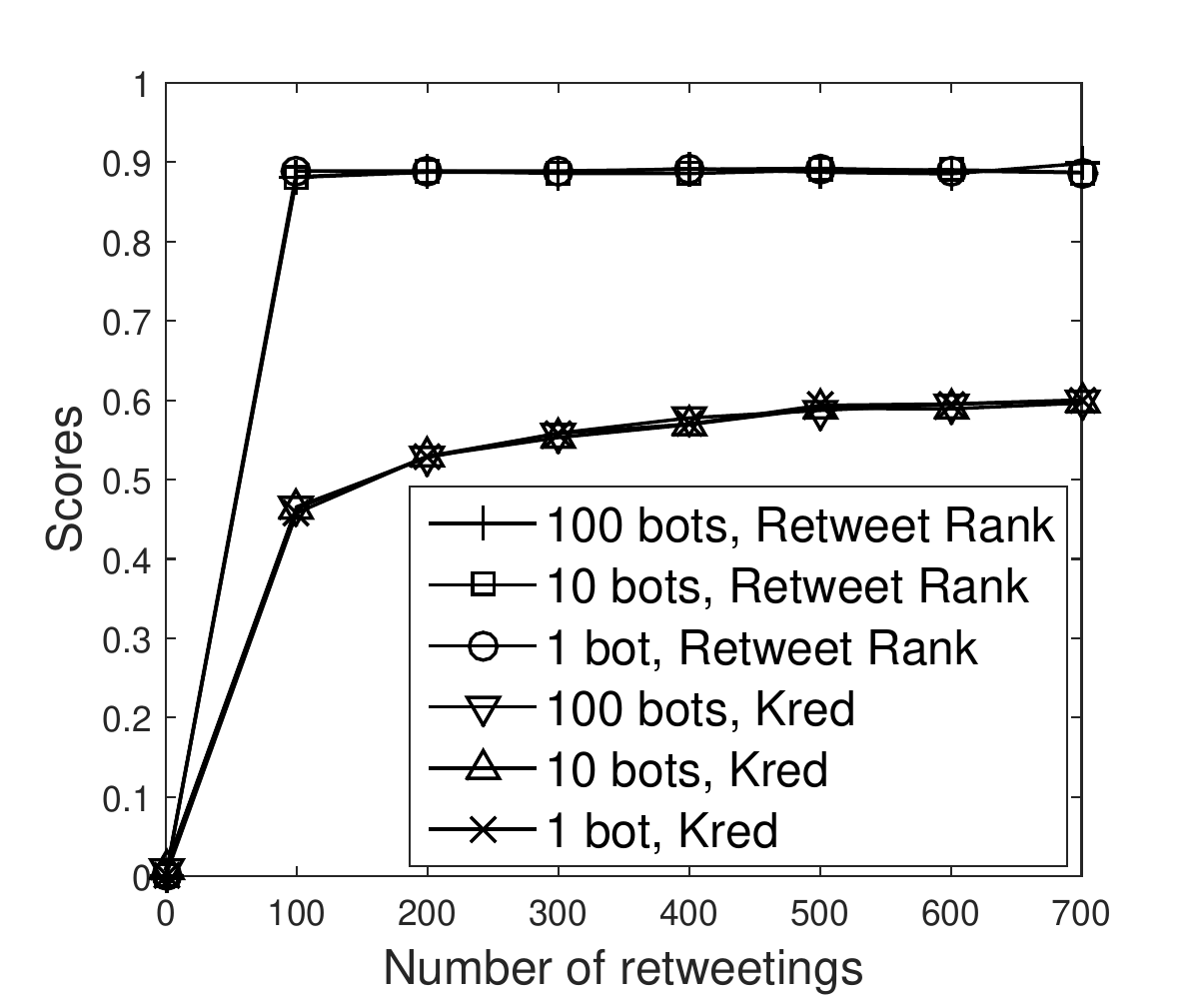}}\hfill
\caption{Manipulation by social botnets with different audience sizes.}
\label{fig:di-audi}
\end{minipage}
\hspace{0.1cm}  
\begin{minipage}{0.48\textwidth}
\centering
\subfigure[Klout Scores]{\label{fig:di-tweet-klout}\includegraphics[width=0.5\textwidth]{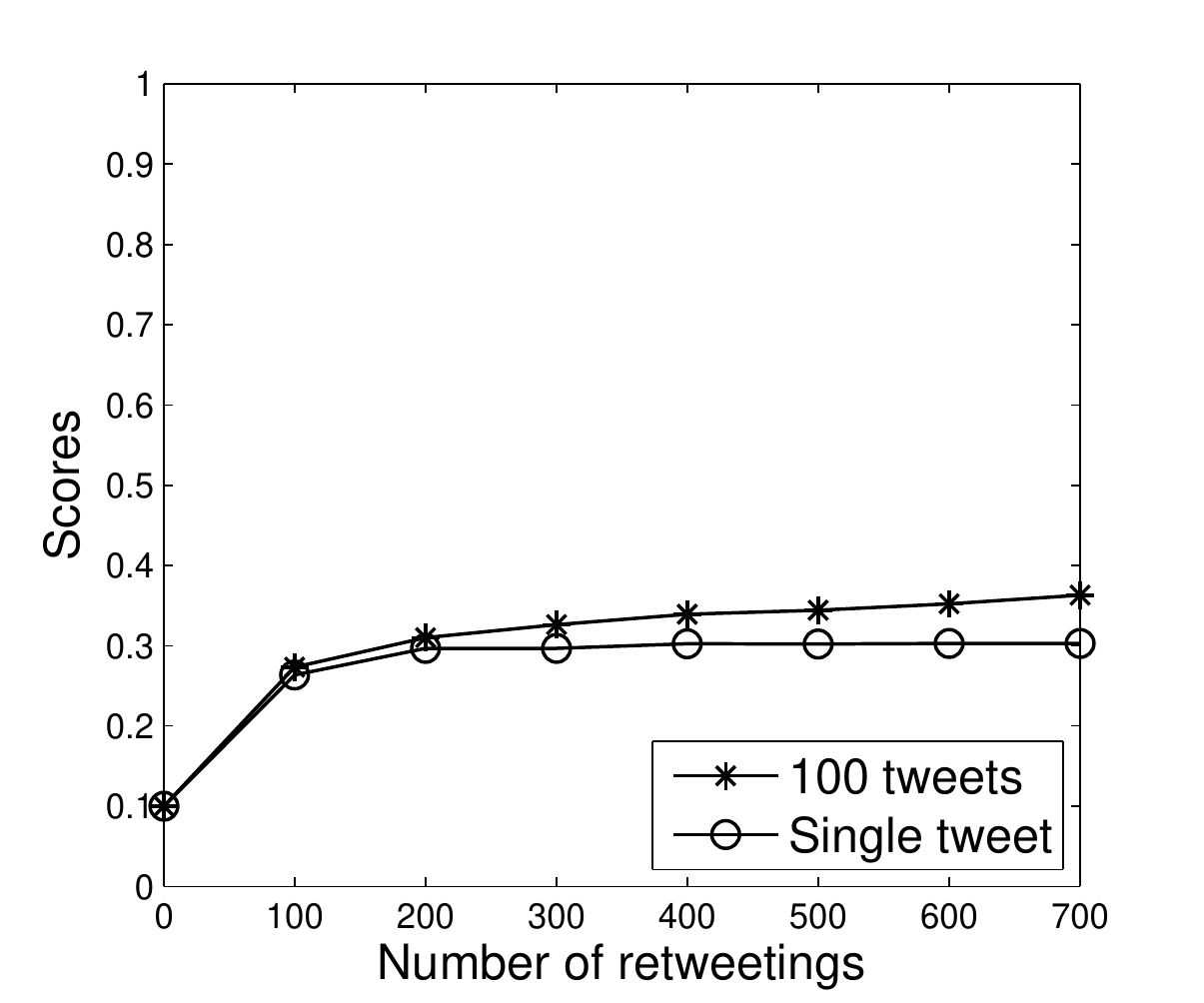}}\hfill
\subfigure[Kred and Retweet Rank Scores]{\label{fig:di-tweet-kred}\includegraphics[width=0.5\textwidth]{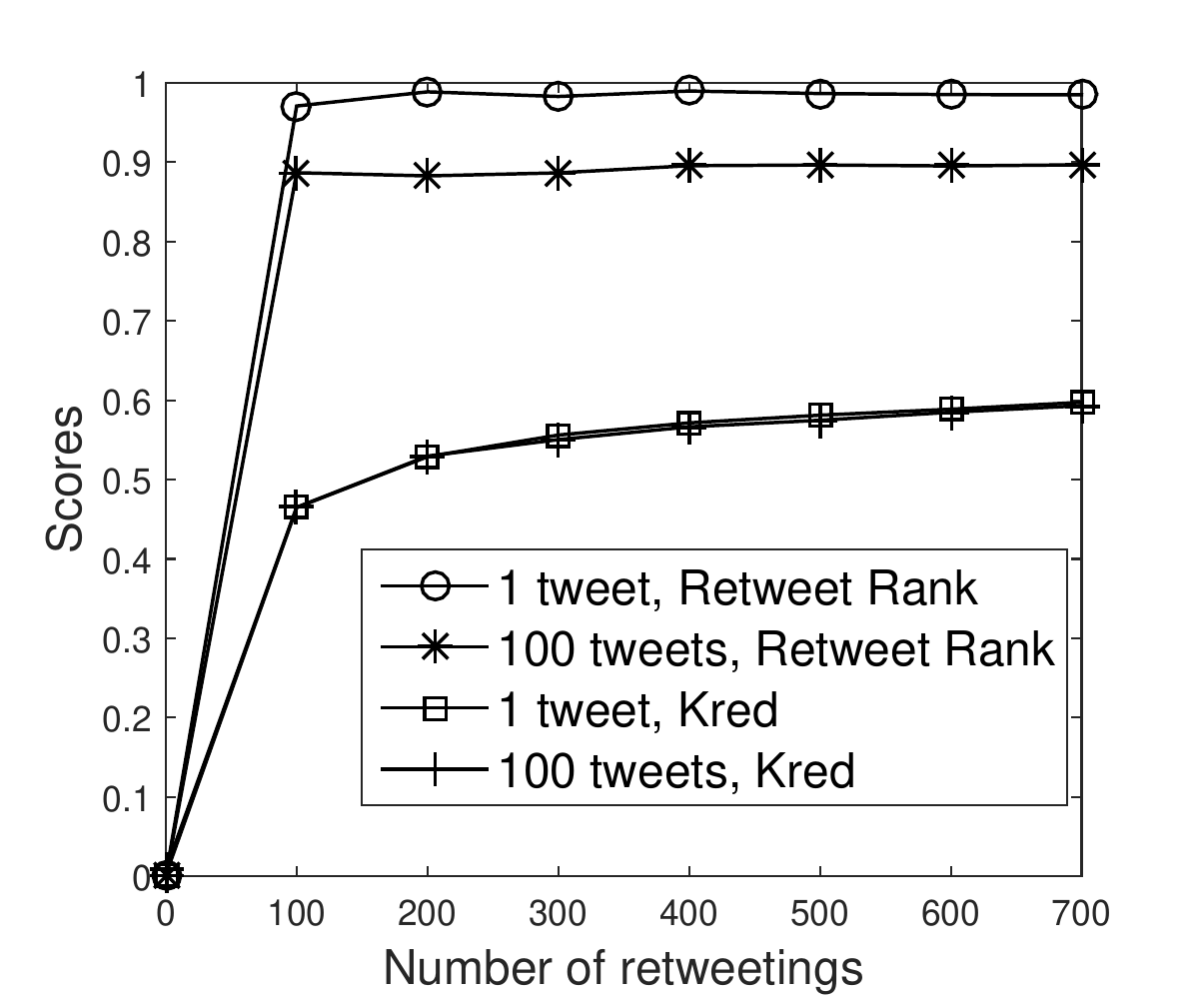}}\hfill
\caption{Manipulation by acting on different number of tweets.}
\label{fig:di-tweet}
\end{minipage}
\end{figure*}

\subsubsection{Impact of Different Actions}
Since almost all digital-influence tools measure a Twitter users digital influence as his ability to drive others to actions, our first experiment aims to evaluate the social botnet's impact with following, retweeting, and mentioning. We do not consider replying, as replying is treated by Twitter as a special type of mentioning and is thus expected to have the same effect as mentioning.

The first set of experiments involves $n=1,000$ social bots, each of which has no interaction with any other Twitter account and hence is not scored by Klout, Kred, or Retweet Rank. Note that Klout assigns an influence score of 10 to new users, while Kred and Retweet Rank both assign a zero score to new users. We randomly choose three disjoint groups, each containing 10 social bots and performing a unique action. For every  social bot in the \emph{following} group, we add 10 randomly chosen social bots as followers each day of the first 10 days and then 100 followers each day of the next 10 days. Likewise, every social bot in the \emph{retweeting} (or \emph{mentioning}) group is retweeted (or mentioned) by randomly chosen social bots 10, 100 and 1000 times each day in the first, second, and the last 10 days, respectively. Since different vendors have different schedules for score updating, we report the social bots' influence scores observed at every midnight. In addition, since the three vendors have different score scales, we normalize different influence scores with respect to the corresponding maximum scores to facilitate direct comparison. In particular, we show $x/100$ and $y/1000$ for a Klout score $x$ and a Kred score $y$, respectively, and report the percentile score for Tweet Rank.

\begin{figure}
\centering
\subfigure[Impact of following]{\label{fig:di-followers}\includegraphics[width=0.24\textwidth]{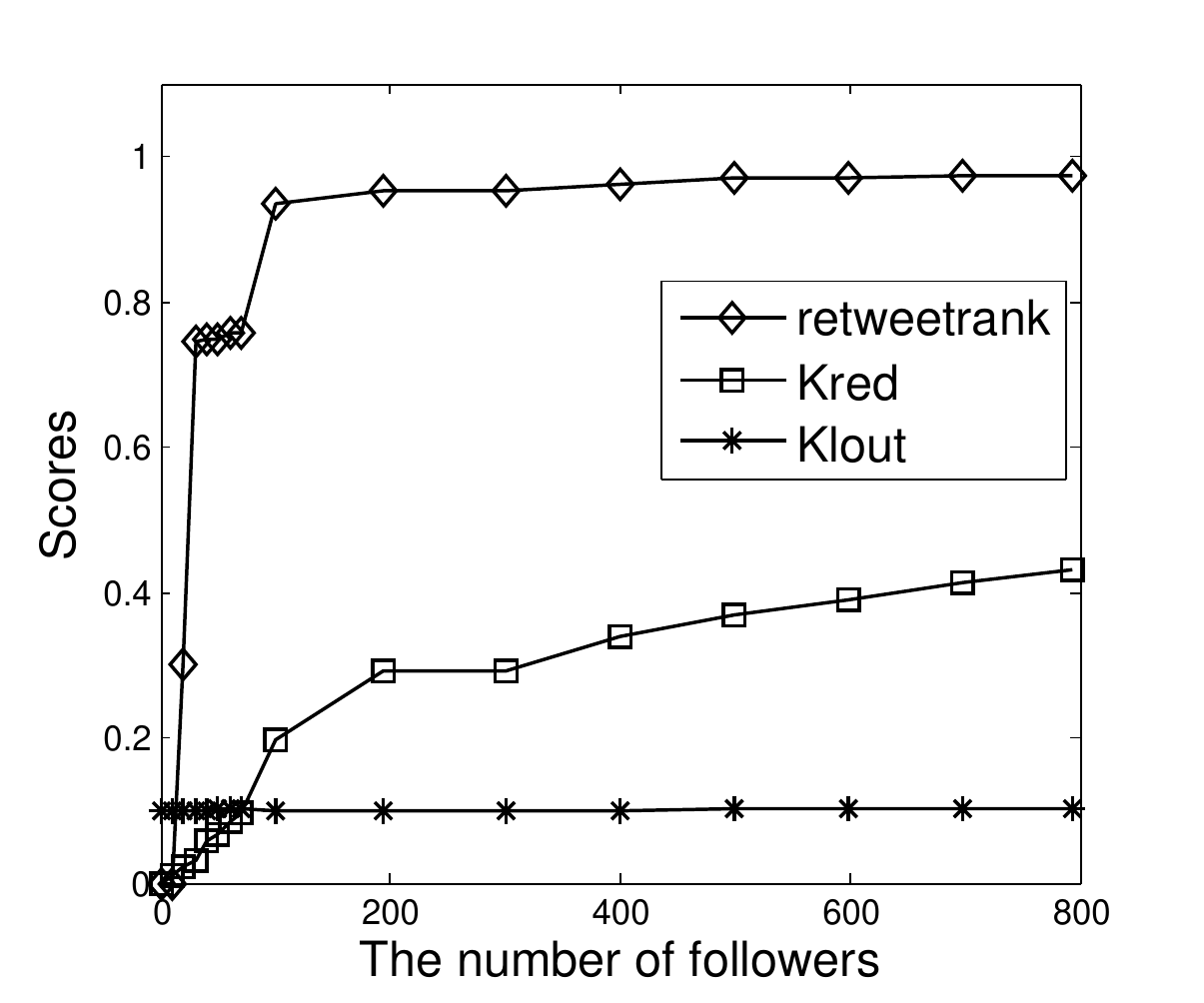}}\hfill
\subfigure[Impact of retweeting]{\label{fig:di-rt}\includegraphics[width=0.24\textwidth]{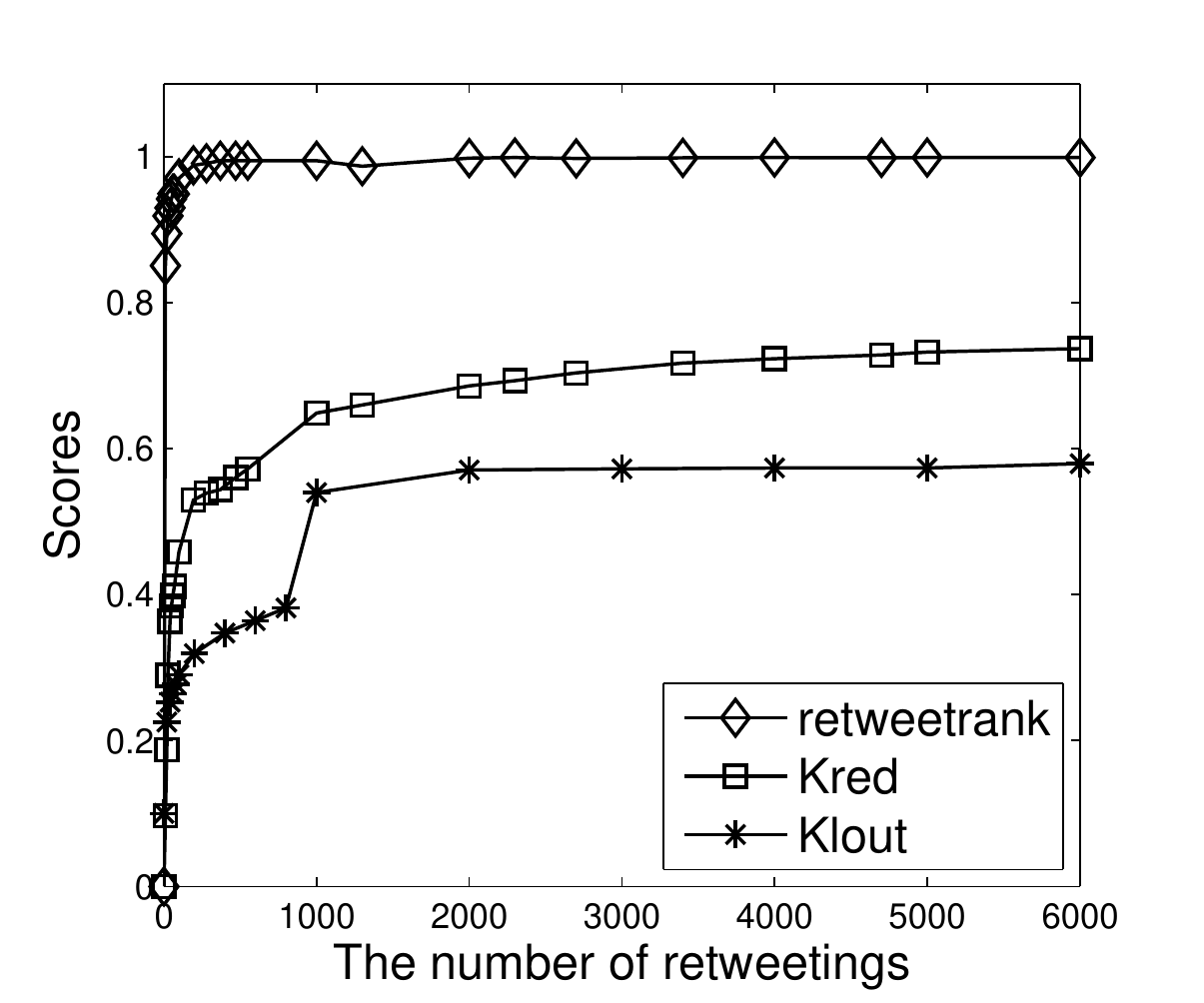}}\hfill

\caption{Manipulating digital influence by following and retweeting.}
\label{fig:di-types}
\end{figure}

Figs.~\ref{fig:di-followers}$\sim$\ref{fig:di-rt} show the impact of following and retweeting\footnote{Mentioning action has similar result with retweeting \cite{ZhangOn13}. We omitted here due to space constraints.} actions on Klout, Kred, and Retweet Rank influence scores, where every data point is the average across the same group. We can see from Fig.~\ref{fig:di-followers} that the Klout influence score is not affected by the number of followers, while both Kred and Retweet Rank influence scores increase as the number of followers increases. This indicates that a social botnet can easily boost the Kred and Retweet Rank influence scores of its members by purposely following each other. Moreover, we can see from Fig.~\ref{fig:di-rt} that all three types of influence scores increase as the number that a user is retweeted increases. On the one hand, this makes much sense, as the higher the frequency in which a user is retweeted, the higher influence of that user has in its local neighborhood. On the other hand, this also renders the influence score measurement system vulnerable to social botnets, as colluding social bots can fake arbitrarily high retweeting frequency for any target user. 

We can also see from Figs.~\ref{fig:di-followers} and \ref{fig:di-rt} that none of our experiments has been able to escalate a user's influence score to an extremely high value, and we conjecture that there are two reasons for such difficulty. First, the audience size is limited, as we only have 1,000 social bots for experiment. We will show in the next set of experiments that the audience size has a large impact on the digital-influence scores. Second, it is likely that all the vendors have set up rules such that it is extremely difficult to achieve almost full digital-influence scores \cite{Klout, Kred, retweetrank}. Nevertheless, at the end of the experiments, Table~\ref{tbl:vendors} shows that all the digital-influence scores being manipulated have exceeded the 90-th percentile.

\subsubsection{Impact of Different Audience Sizes}
In this set of experiments, we measure the impact of different audience sizes on digital-influence scores.  Recall that we define a user $u$'s \emph{audiences} as the set of users who have retweeted, mentioned, or followed $u$. We then try to answer the following questions. First, for two users with different audience sizes but the same numbers of actions, will the one with more audiences have a higher digital-influence score than the other? Second, it there an upper limit for a single social bot to manipulate the influence score of a target user? There could be two alternative answers to each of the two questions. On the one hand, if the digital-influence scores are related to both the number of incoming actions and the audience size, a single user should have limited power to manipulate the target user's influence score, and we thus need a large social botnet to manipulate the target user's influence score to some extremely high value. On the other hand, if the digital-influence scores may not relate to the audience size, then 100 incoming actions from 100 different social bots would yield the same result as 100 incoming actions from a single social bot.

To verify which of the two conjectures is correct, we build three social botnets with 1, 10, and 100 bots, respectively. For each social botnet, we set 10 target users and retweet 100 tweets of each target each day for seven days. In other words, each bot in the three social botnets  retweets 100, 10, and 1 times per day, respectively.  Fig~\ref{fig:di-audi} shows the digital-influence scores of the 30 targets. As we can see, the audience size has no impact on both Kred and Retweet Rank scores but has large impact on the Klout scores. Specifically, the larger the audience size, the higher the Klout scores, and vice versa. Moreover, the Klout scores experience a sharp increase in the first day and then increase much slower in the following days. As a result, a single social bot can manipulate a target user's Kred and Retweet Rank scores with a large number of actions, while both large audience sizes and significant actions are necessary to obtain high Klout scores. We also conclude that Klout is more resilient to the social botnet than Kred and Retweet Rank are.

\subsubsection{Impact of Tweet Popularity}
From the first set of experiments, we can see that the retweeting is the most effective way to manipulate a target user's digital-influence score. Given the same audience, the attacker can either retweet a single tweet of the target user to make this tweet very popular or retweet many of target user tweets so that each tweet will be less popular. We would like to answer which strategy will yield higher digital-influence score, i.e., whether tweet popularity has any impact on the digital-influence scores? To answer this question, we build a social botnet with 100 bots. We then select two groups with each containing 10 target users for manipulation. For the first group, each target first publishes a tweet, which is then retweeted by all the 100 social bots; while for the second group, each target user publishes 100 tweets, each of which is retweeted by one social bot. We repeat this process daily and measure the digital-influence scores of the 20 targets. As we can see from Fig~\ref{fig:di-tweet}, the tweet popularity has no impact on both Klout and Kred scores and limited impact on the Retweet Rank score. In particular, adopting the second strategy will lead to a slightly higher Retweet Rank score of the target user.

\subsubsection{Speed of Digital-influence Manipulation}
Our last experiment targets evaluating how fast influence scores can be manipulated. Same with the last set of experiments, we choose the retweeting as the action due to its effectiveness. For this purpose, we randomly select another three different groups of social bots, each containing 10 target bots. Every bot in the first, second, and third groups is retweeted 10, 100, and 1,000 times every day by random bots until the scores reach the 90th percentile, which corresponds to 50, 656, and 90 in Klout, Kred, and Retweet Rank, respectively.

\begin{figure}[t]
\centering
\subfigure[80-th Percentile]{\label{fig:di-speed80}\includegraphics[width=0.24\textwidth]{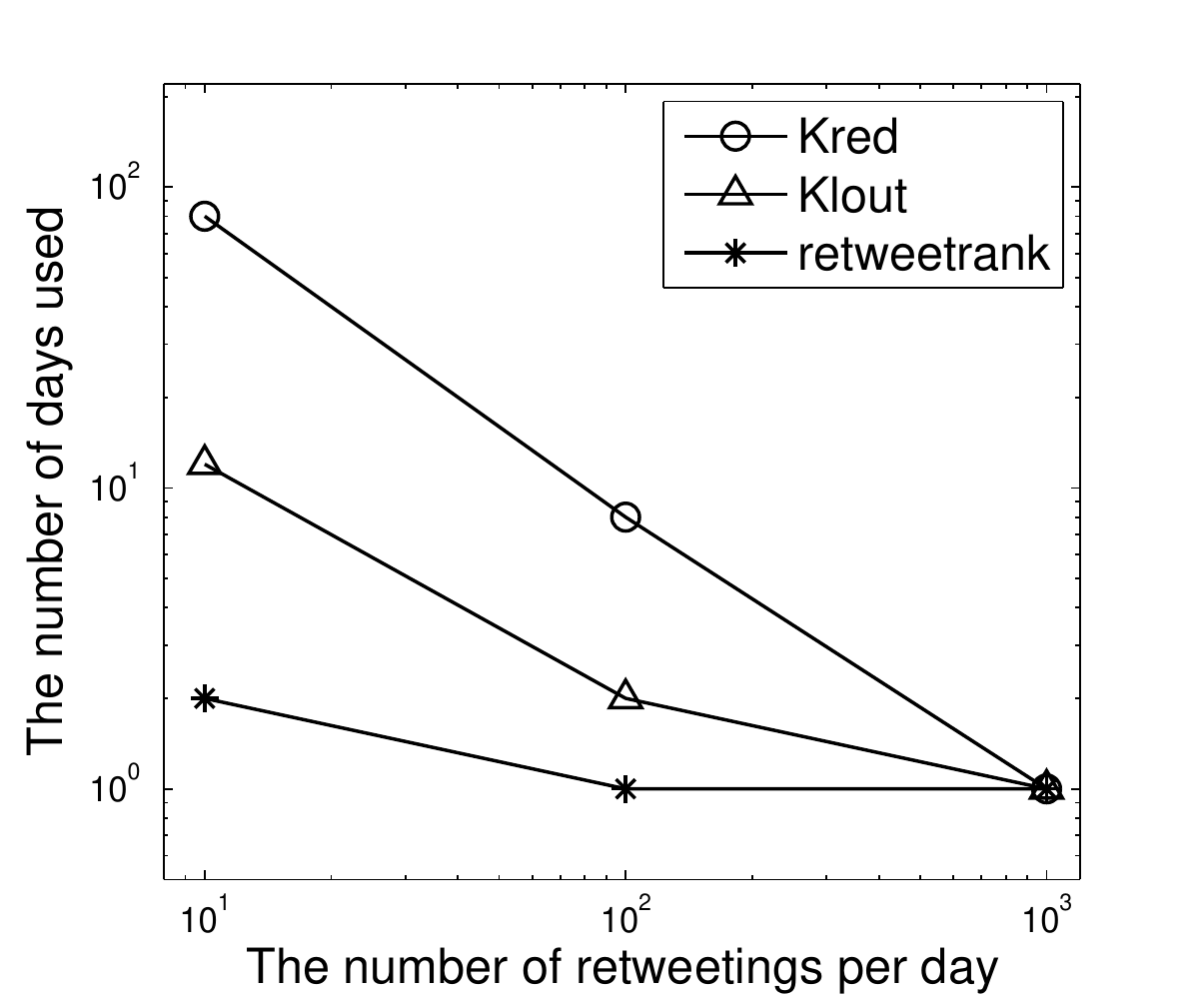}}\hfill
\subfigure[90-th Percentile]{\label{fig:di-speed90}\includegraphics[width=0.24\textwidth]{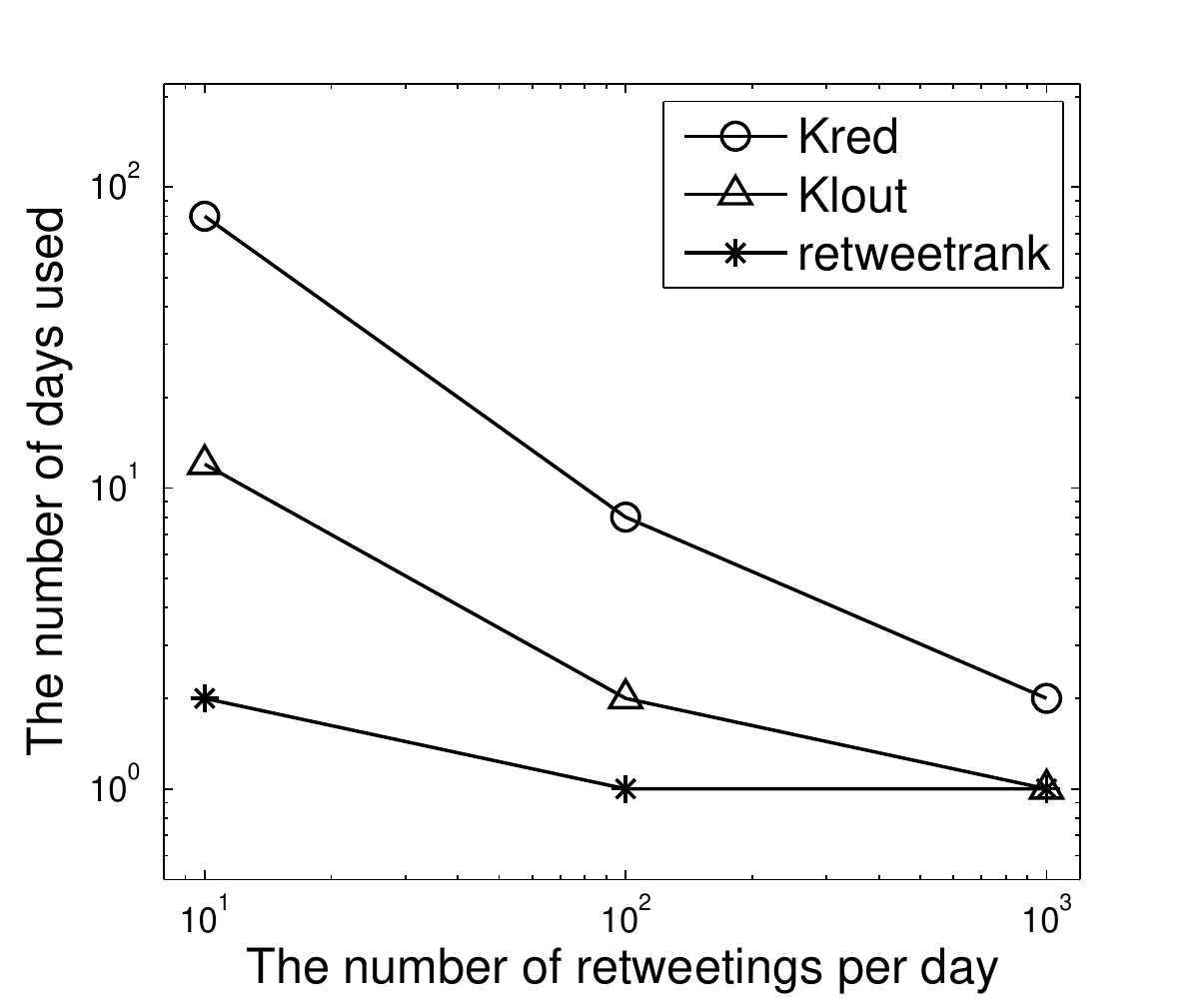}}\hfill
\caption{Under different retweeting speeds, the number of days needed to manipulate digital influence scores from nothing into 80-th and 90-th percentiles.}
\label{fig:di-speed}
\vspace{-.2in}
\end{figure}

Fig.~\ref{fig:di-speed} further shows the impact of retweeting frequency on the change of influence scores, where every data point represents the average across the same bot group. In particular, we can see that the number of days needed to increase the group average influence score from the initial value to 80-th or 90-th percentile is approximately inversely proportional to the retweeting frequency. In addition, we can see that for all three vendors, it is possible by retweeting 1000 times per day to reach the 80-th percentile and the 90-th percentile with only one and two days, respectively.

\begin{figure}
\centering
\subfigure[True Positive rate]{\label{fig:df1-gamma-tp}\includegraphics[width=0.24\textwidth]{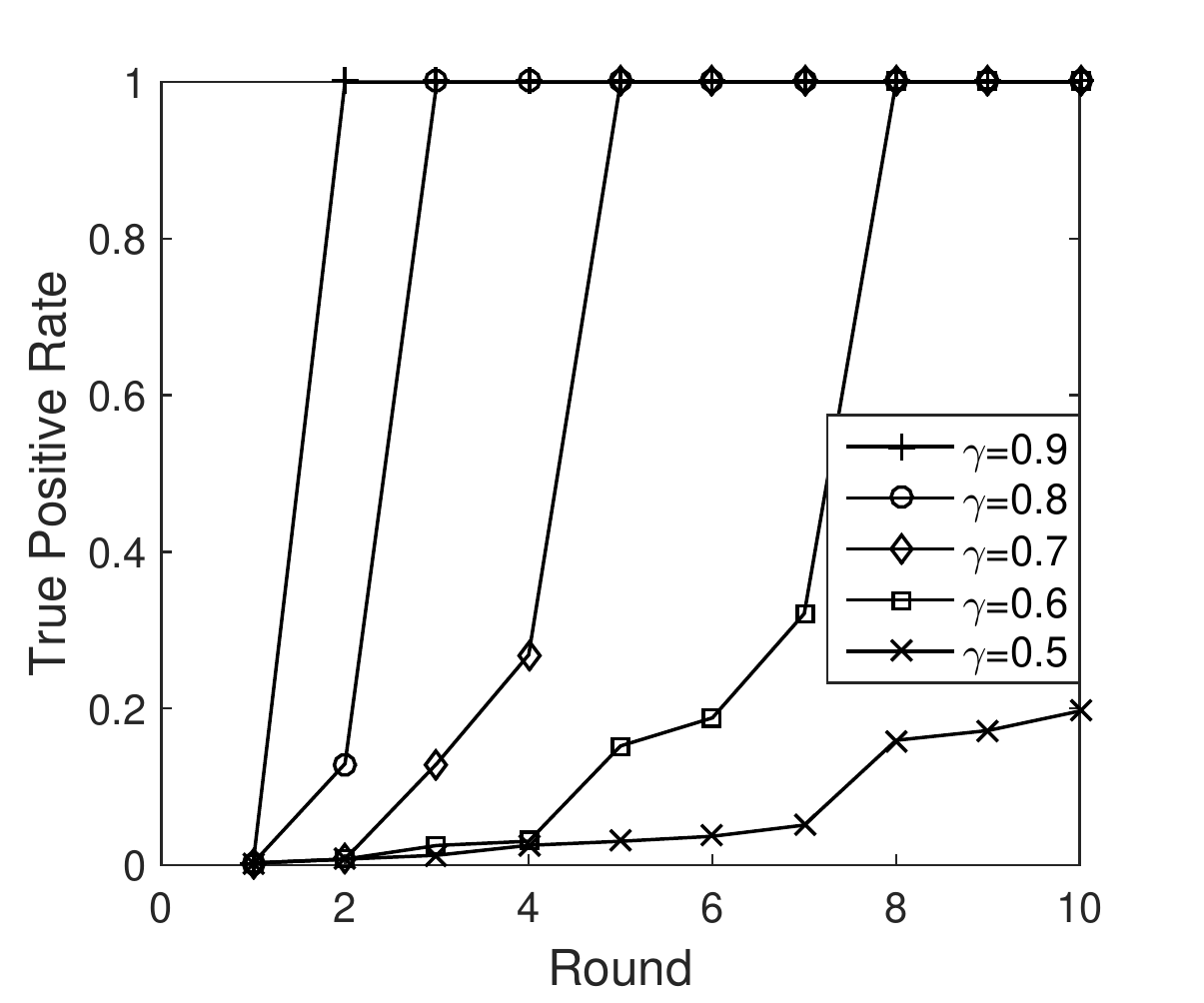}}\hfill
\subfigure[False Positive rate]{\label{fig:df1-gamma-fp}\includegraphics[width=0.24\textwidth]{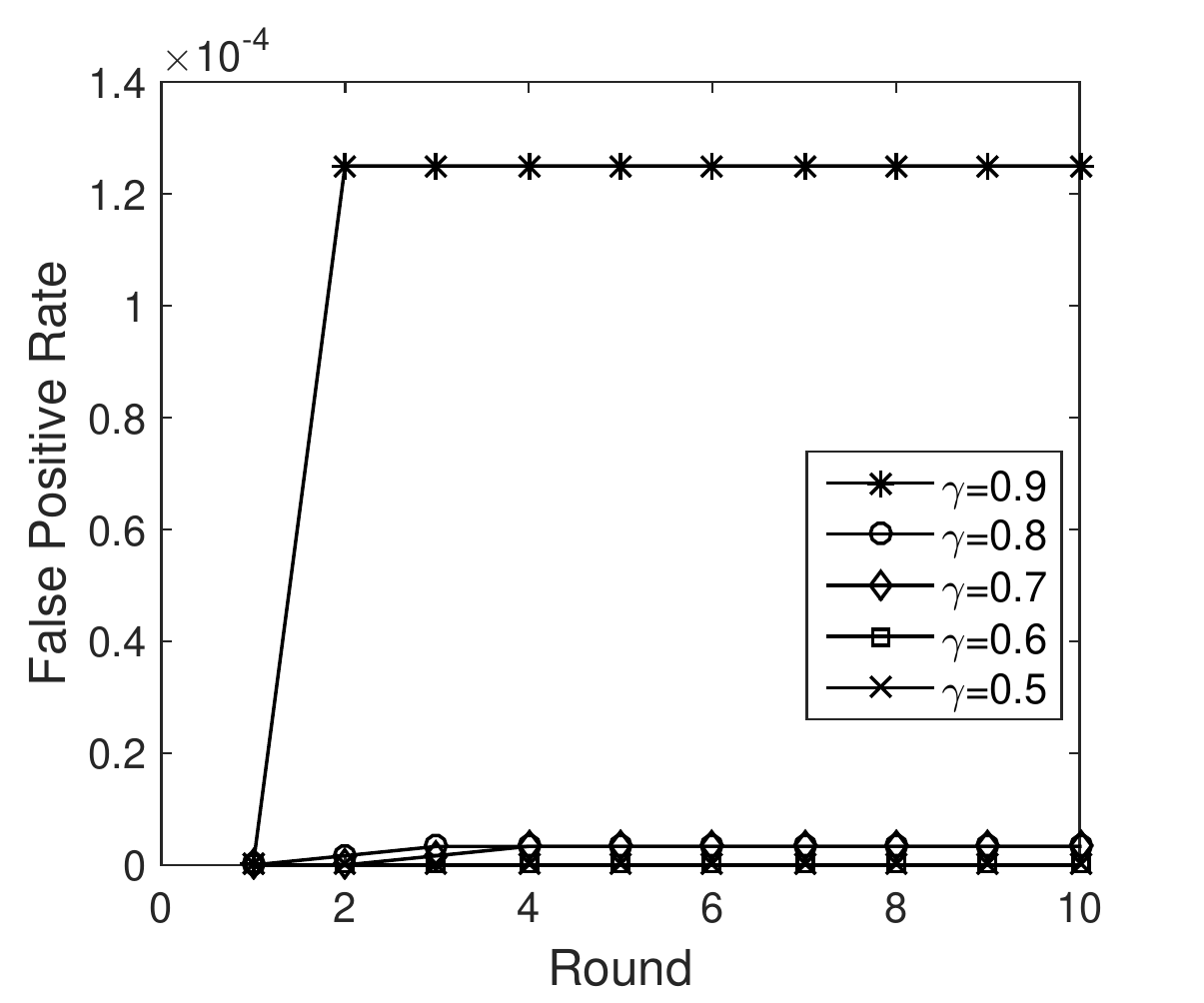}}\hfill

\caption{The true and false positive rates with different $\gamma$s.}
\label{fig:df1-gamma}
\vspace{-.2in}
\end{figure}

\begin{figure*}
\centering
\subfigure[True positive rate (Recall)]{\label{fig:df1-compare-tp}\includegraphics[width=0.24\textwidth]{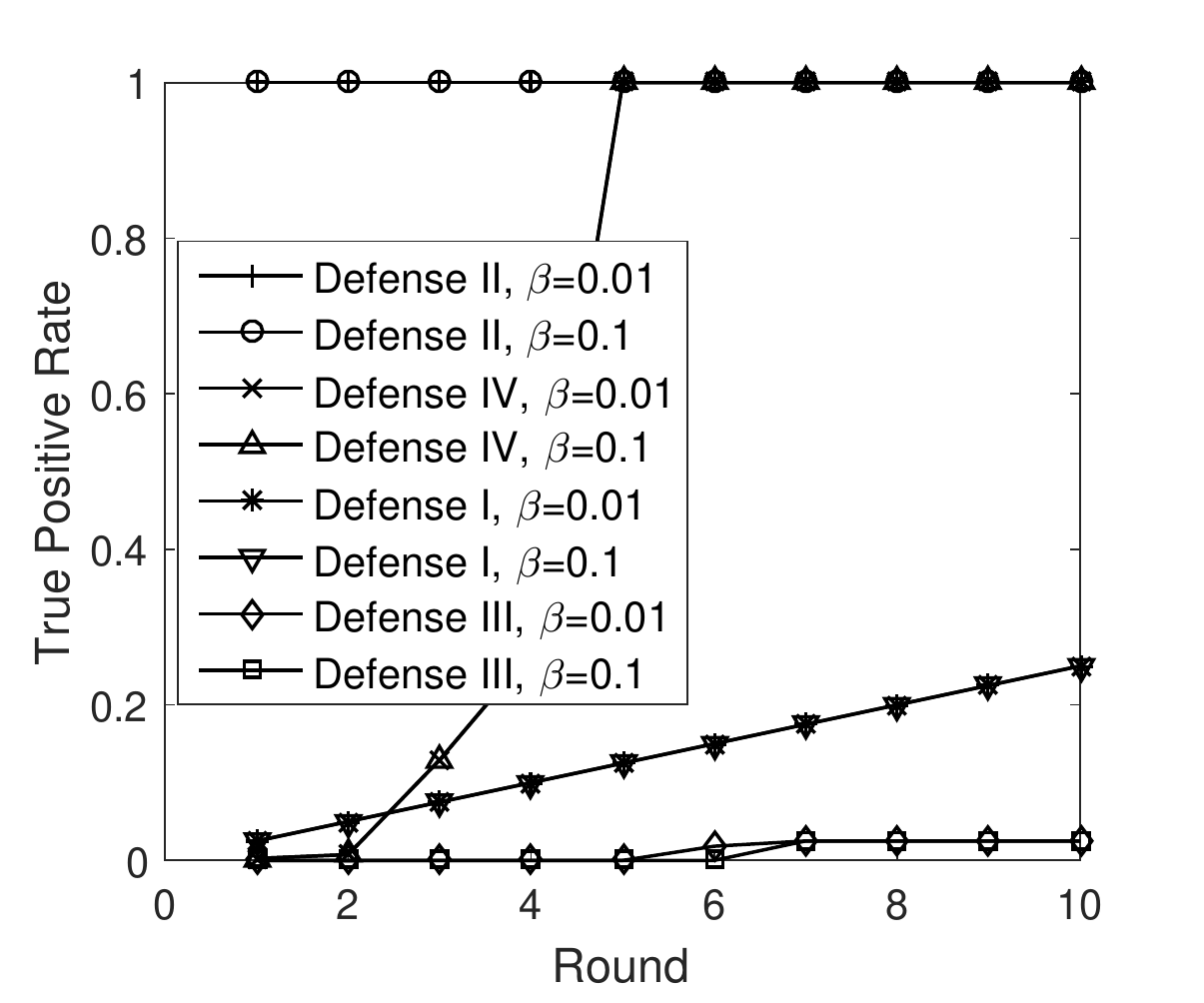}}\hfill
\subfigure[False positive rate]{\label{fig:df1-compare-fp}\includegraphics[width=0.24\textwidth]{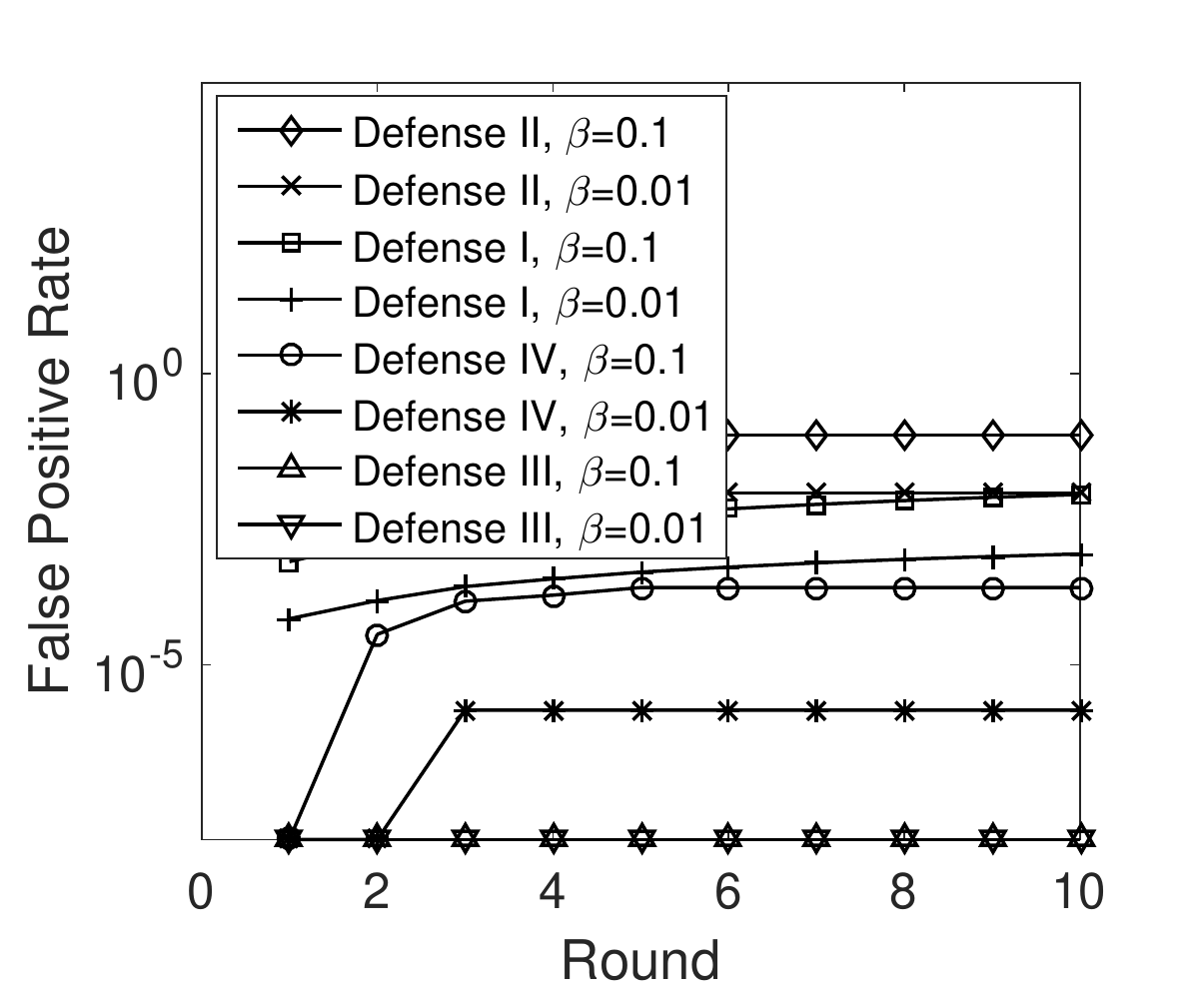}}\hfill
\subfigure[Precision]{\label{fig:df1-compare-precision}\includegraphics[width=0.24\textwidth]{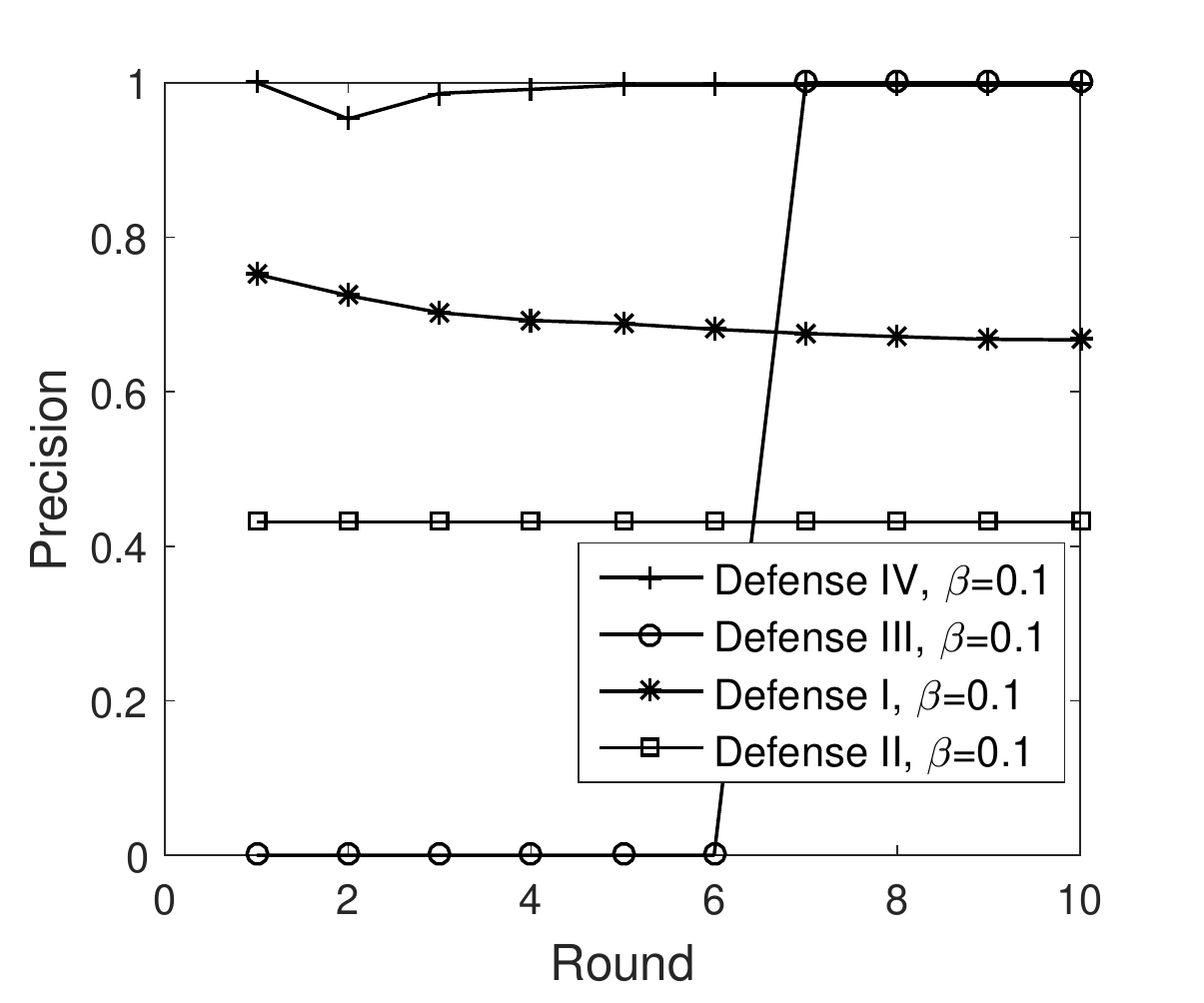}}\hfill
\subfigure[Overall performance]{\label{fig:df1-compare-overall}\includegraphics[width=0.24\textwidth]{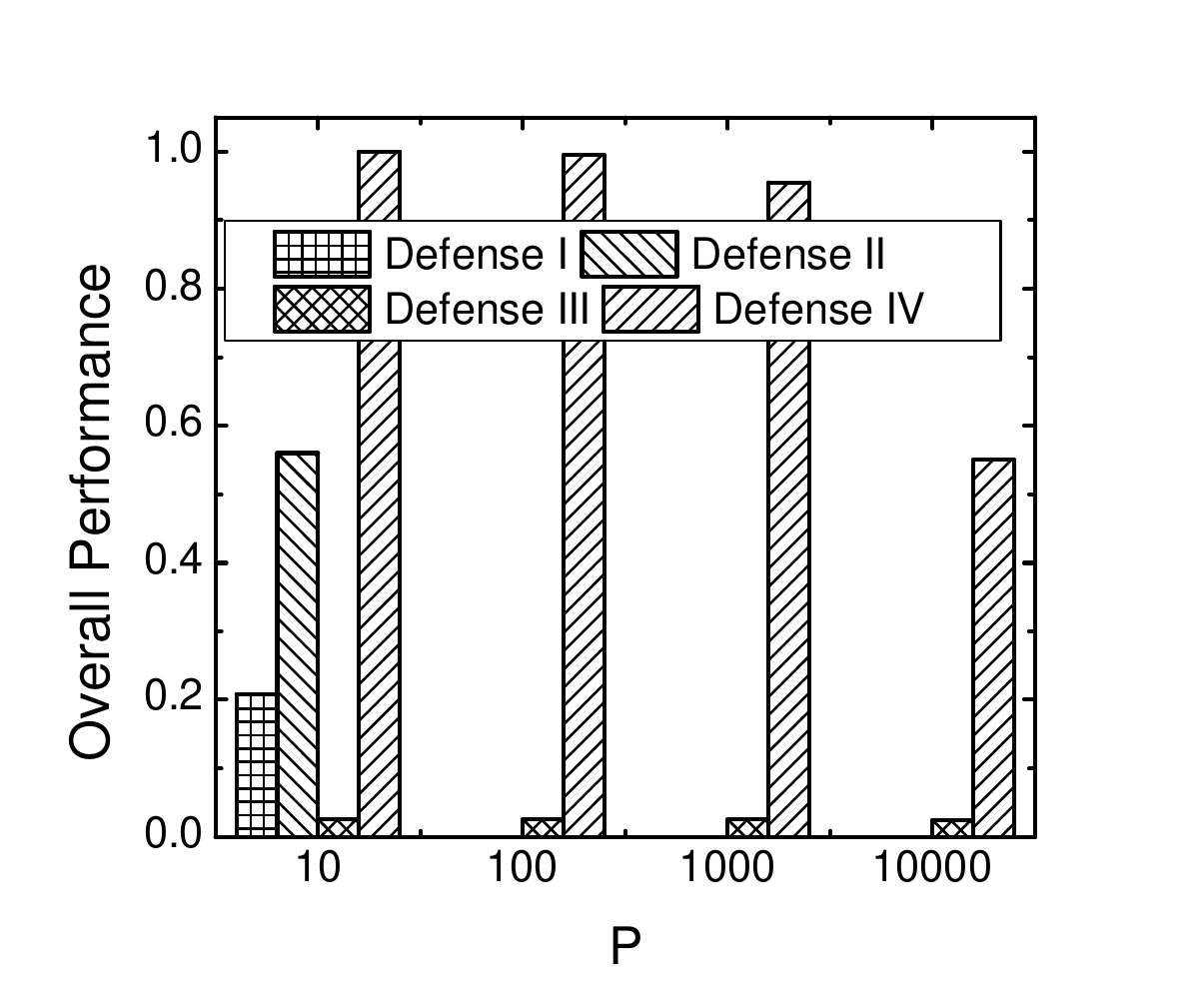}}\hfill
\caption{Performance Comparison.}
\label{fig:df1-compare}
\end{figure*}

\subsubsection{Remarks}

We have two remarks to make.

First, at the end of the experiments, no account was suspended by Twitter, as none of the accounts had conducted illegitimate activities such as spamming and aggressively following and unfollowing which could trigger suspension by Twitter.

In addition, both the spam distribution and the digital influence manipulation attacks show that social
botnet has significant advantage over isolated bots. In particular, even though the attacker can
disseminate spams to the same legitimate recipients using isolated bots, those isolated bots will
be quickly suspended under Twitter's current policy or its variation and thus incurs significant
cost for the attacker. Moreover, it is extremely difficult for the attacker to boost a single bot's
digital influence score solely by letting the target bot tweet or retweet other legitimate users'
tweets, as long as the digital influence score is not entirely determined by user's outgoing
interactions, as it is difficult for bots to attract incoming interactions from legitimate users.

\section{Defenses}\label{sec:defend}
In this section, we propose two countermeasures for the two attacks above, respectively.

\subsection{Defense against Botnet-based Spam Distribution}
Recall that in social botnet-based spam distribution, the attacker exploits retweeting trees to distribute spams (i.e., tweets with malicious URLs) such that only the bots on the first $M$ (e.g., $M=1$ in Twitter currently) levels of retweeting trees will be suspended.

To defend against this attack, we propose to track each user's history of participating in spam distribution and suspend a user if his accumulated suspicious behaviors exceed some threshold. Specifically, for each user $v$ we maintain a \emph{spam score} $s_v$, which is updated every time user $v$ retweets a spam. Once $s_v$ exceeds a predefined threshold, user $v$ is labeled as a spammer and suspended.

We now discuss how $s_v$ is updated. Our intuition is that the closer the user to the spam source, the more likely he is a member of the social botnet. The reason is that social botnet usually prefers shorter retweeting path for fast spam dissemination, while a spam tweet traversing a long retweeting path, i.e., sequentially retweeted by many accounts, incurs large delay and gives more time for Twitter to detect them. To reflect this idea, whenever a user retweets a spam, we update his spam score as
\begin{equation}
s_v = s_v+\gamma^d\;,
\label{eq:spamscore}
\end{equation}
where $d$ is the number of retweeting hops between the spam source to $v$, and $\gamma\leq 1$ is the attenuation factor of distance. Note that $d=0$ if user $v$ is the source of a spam. In this way, any user who has retweeted  spams is punished by having his spam score increased. Once a user's spam score exceeds certain predetermined threshold, the user is suspended. Note that Twitter currently suspends a user whenever he has published one spam tweet. To mimic the current Twitter policy, we can set the threshold as one.

\subsubsection{Evaluation}
Similar to Section~\ref{sec:evaluationNSB}, we built a Twitter subnet composed of 6000 legitimate users and 400 social bots. Each social bot has $|\mathcal{F}_i|$ legitimate followers, where $|\mathcal{F}_i|$ is drawn from Gaussian distribution with $\mu=32$ and $\delta^2=5$, and each social bot follows all other social bots. We assume that the attacker builds the optimal retweeting tree according to Section~\ref{sec:heusolution}. We adopt similar parameters as in Section~\ref{sec:evaluationNSB} by setting $M=3$, $\alpha=0.2$, $c=10$, and $K=10$.

To model the spam campaign, we conduct the experiment in multiple rounds. In each round, we randomly select one social bot to publish a spam tweet and other social bots that have not been suspended to retweet the spam tweet according to the retweeting forest. We assume that the legitimate users retweet the spam occasionally with probability $\beta$. At the end of each round, we update the spam scores for each bot and legitimate user according to Eq.~(\ref{eq:spamscore}). If $s_v\geq 1$ for some user $v$, we remove $v$ from the simulated network. We then start the next round by randomly selecting a bot as the new source and continue the spam distribution. To reduce the randomness, we repeat each experiment 100 times and report the average results.

We evaluate the proposed defense in comparison with three other baseline defenses as follows.
\begin{itemize}
\item \textbf{Defense I}. The original defense that suspends the users in the first $M$ levels from the spam source, which is considered by the attacker to launch the optimal spam distribution as discussed in Section 3.2.
\item \textbf{Defense II}. Any user who retweets $\delta$ spams is suspended, regardless of its distance to the spam source, where $\delta$ is a system parameter.
\item \textbf{Defense III}. Assume that user $v$ has retweeted a spam tweet $t$ within $M$ hops from the source and that $t$ has been retweeted by $n_t$ users in total. The spam score for $v$ is updated as
\[
s_v = s_v + 1/\log(1+n_t)\;.
\]
Defense III extends Defense I and II by taking the popularity of individual spam tweet into account. The intuition is that the more users retweet a spam tweet, the more deceiving the spam tweet, and the less likely that any individual user who retweets it is a bot. We also use the system parameter $\delta$ as the suspending threshold for this defense scheme.
\item \textbf{Defense IV}. The proposed defense.
\end{itemize}

We use four metrics to compare the proposed defense with the three baseline defenses. Let $N_{\text{bots}}$ and $N_{\text{legit}}$ be the numbers of social bots and legitimate users, respectively. Also let $S_{\text{bots}}$ and $S_{\text{legit}}$ be the numbers of suspended social bots and legitimate users, respectively. We define the following four metrics.
\begin{itemize}
\item \textbf{True positive rate ($\mathtt{TPR}$):} the ratio of the suspended bots over the total bots, which can be computed as $S_{\text{bots}} / N_{\text{bots}}$. This metric is also referred to as recall.
\item \textbf{False positive rate ($\mathtt{FPR}$):} the ratio of suspended legitimate users over all the legitimate users, which can be computed as $S_{\text{legit}} / N_{\text{legit}}$.
\item \textbf{Precision:} the ratio of  suspended bots over all suspended users, which can be computed as $S_{\text{bots}}/(S_{\text{legit}} + S_{\text{bots}})$.
\item \textbf{Overall performance:} $\mathtt{TPR} - P\cdot\mathtt{FPR}$ where $P$ is the penalty parameter for $\mathtt{FPR}$.
\end{itemize}
We adopted overall performance here because $\mathtt{FPR}$ has much larger impact than $\mathtt{TPR}$ from the service provider's point of view, as suspending a legitimate user damages its reputation and incurs severer consequences than a social bot evading detection.

Fig.~\ref{fig:df1-gamma} shows the true positive and false positive rates with different $\gamma$s which are the attenuation factors of distance. As we can see, the proposed defense could quickly detect all the social bots when $\gamma$ is large. Specifically, when $\gamma=0.9$, all the social bots can be detected and suspended in the second round. Moreover, as expected, the ratio of true suspension at the same round will decrease as $\gamma$ decreases. Finally, there is an anticipated tradeoff between false and true positive rates for different $\gamma$s. The larger $\gamma$, the higher true positive rate but also the higher false positive rate, and vice versa. In the following experiments, we set $\gamma=0.7$ by default.

Fig.~\ref{fig:df1-compare} compares the proposed defense with the three baselines under different $\beta$s, i.e., the probability of a legitimate user retweeting a spam tweet. We can make five observations here. First, Fig.~\ref{fig:df1-compare-tp} shows that $\beta$ has no impact on the $\mathtt{TPR}$, but Fig.~\ref{fig:df1-compare-fp} shows that the larger $\beta$, the higher $\mathtt{FPR}$ for all four defenses, which is consistent with the definition of $\beta$. Second, Fig.~\ref{fig:df1-compare-tp} shows that \textbf{Defense II} has the highest $\mathtt{TPR}=100\%$ in every round. The reason is that when $\delta=1.0$ all the social bots will be suspended after the first round. \textbf{Defense IV} has the second highest $\mathtt{TPR}$ and needs about five rounds to approach 100\%. \textbf{Defense I} and \textbf{III} both have lower $\mathtt{TPR}$ because they only examine the users in the first $M=3$ levels instead of the whole network. Third, Fig.~\ref{fig:df1-compare-fp} shows that the $\mathtt{FPR}$ of the proposed defense is lower than those of \textbf{Defenses I} and \textbf{II} but slightly higher than that of \textbf{Defense III}. This is because \textbf{Defense I} and \textbf{II} have much less restrict conditions for suspending users than \textbf{Defense IV}. Fourth, Fig.~\ref{fig:df1-compare-precision} shows that \textbf{Defenses IV}  and \textbf{III} have the precision close to 100\% after the first and sixth rounds, respectively, as they both incur low false positives and sufficiently high true positives. \textbf{Defense III} has 0\% of precision before the sixth round because it did not suspend any social bot in the first five rounds. In contrast, both \textbf{Defense I} and \textbf{II} have relatively low precision because they misclassified many normal users as bots. Finally, Fig.~\ref{fig:df1-compare-overall} shows that the proposed defense has better overall performance than all three baselines with penalty factor $P$ varying from 10 to 10000. The reason is that \textbf{Defense IV} has much lower $\mathtt{FPR}$ than \textbf{Defense I} and \textbf{II} and higher $\mathtt{TPR}$ than \textbf{Defense III}. In summary,  the proposed defense outperforms all three baselines as it can effectively detect social bots while being friendly to legitimate users.

\begin{table}[t]
\footnotesize
    \caption{Four action networks for evaluation, where 'F' and 'I' refer to following and interaction, respectively. }
    \centering
    \begin{tabular}{|c|c|c|c|c|}
       \hline
        {Area} & \#Users & \#F-edges & \#I-edges (\#Weights)\\
       \hline
        Tucson (\texttt{TS}) &  28,161 & 830,926 & 162,333 (669,006) \\
        Philadelphia (\texttt{PI}) &  144,033 & 5,462,013 & 1,434,375 (4,972,689)\\
        Chicago (\texttt{CI}) &  318,632 & 14,737,526 & 3,631,469 (12,010,986)\\
        Los Angeles (\texttt{LA}) & 300,148 & 18,333,774 & 4,391,542 (14,048,838)\\
        \hline
    \end{tabular}
    \label{tlb:datasets}
\end{table}

\begin{figure*}
\centering
\begin{minipage}{0.48\textwidth}
\centering
\subfigure[Top-$10$-percent accuracy]{\label{fig:df2-3}\includegraphics[width=0.5\textwidth]{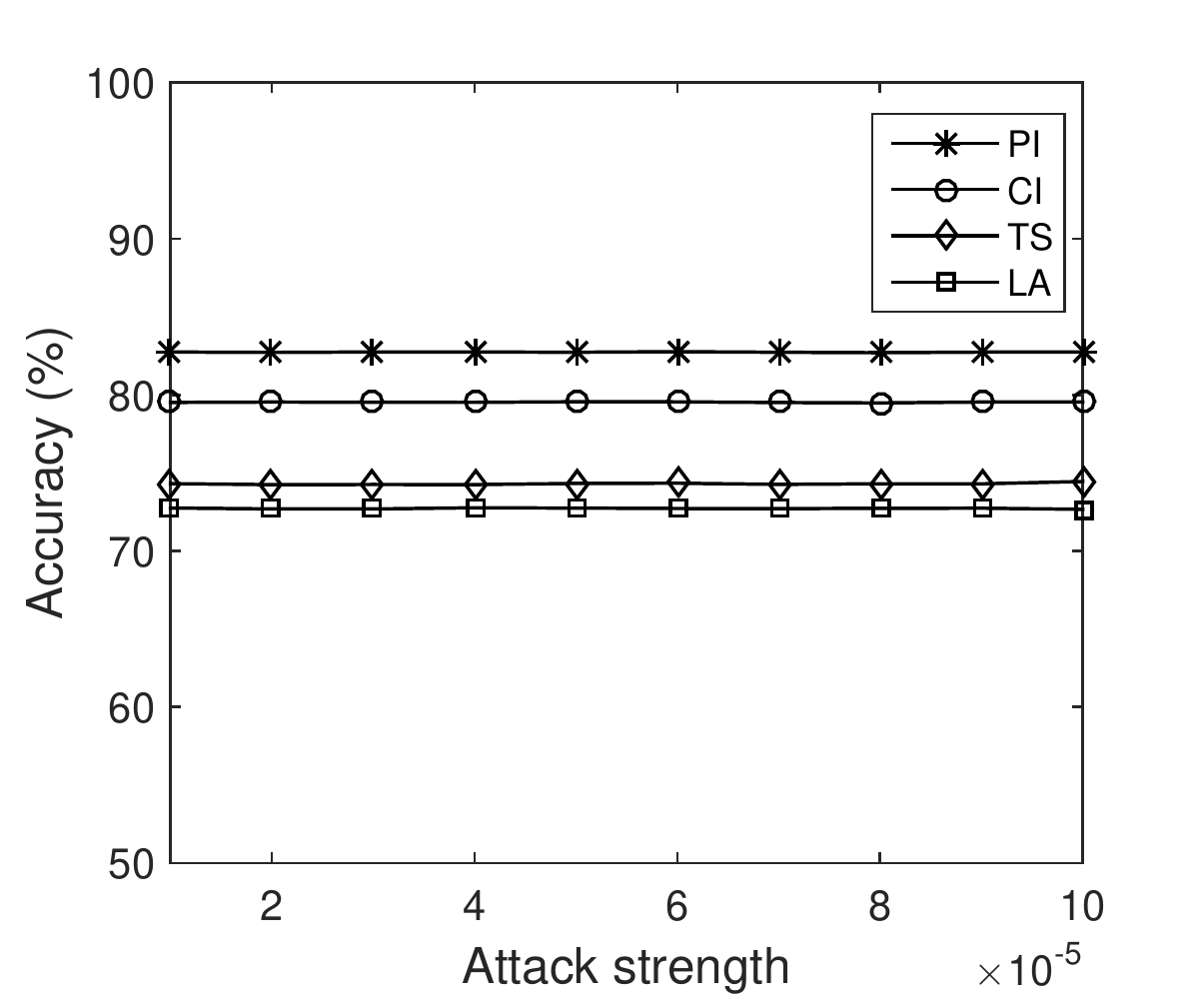}}\hfill
\subfigure[Social-bot influence ranking]{\label{fig:df2-4}\includegraphics[width=0.5\textwidth]{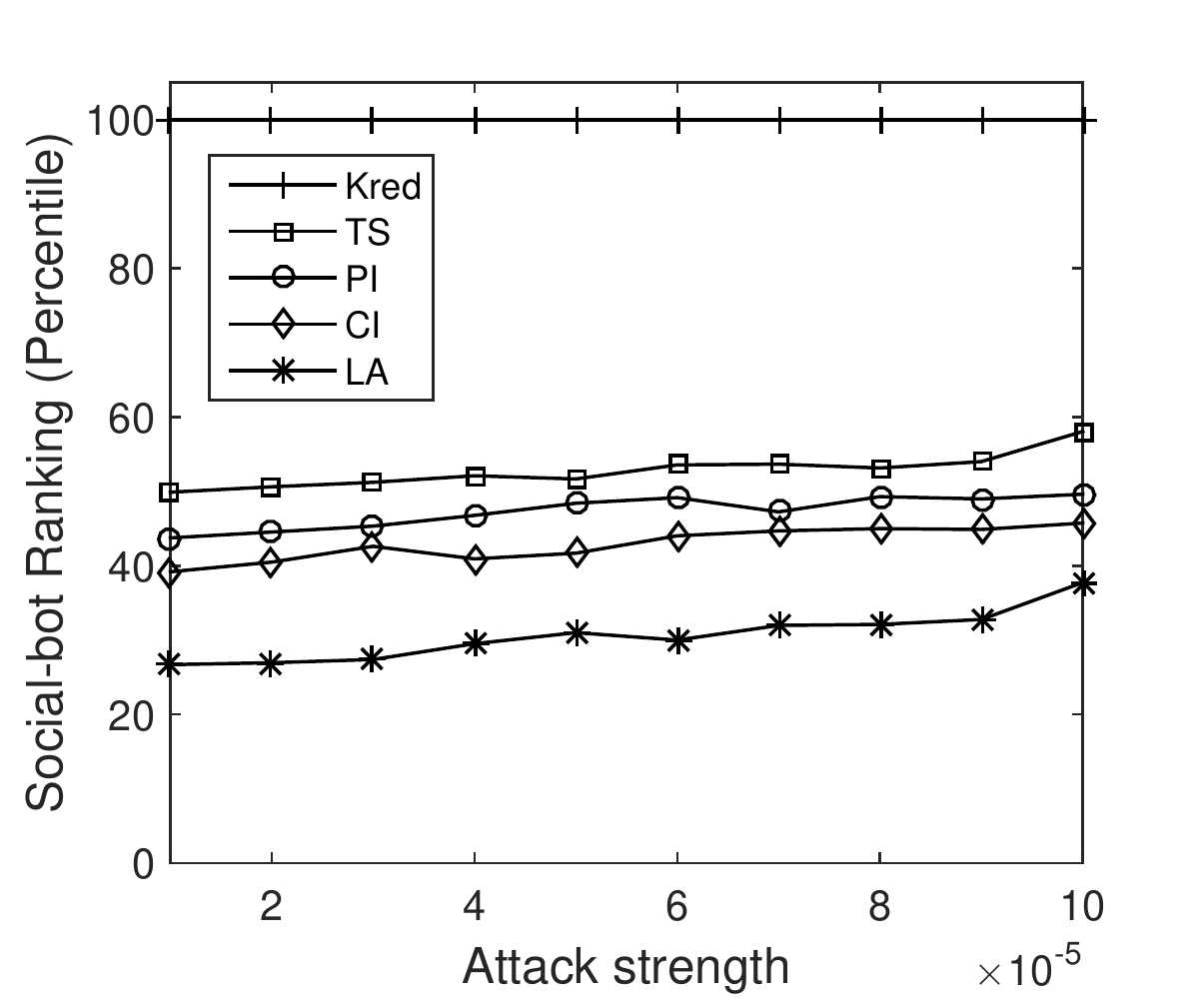}}\hfill
\caption{The performance under the random attack.}
\label{fig:df2-attack1}
\end{minipage}
\hspace{0.1cm}  
\begin{minipage}{0.48\textwidth}
\centering
\subfigure[Top-$10$-percent accuracy]{\label{fig:df2-7}\includegraphics[width=0.5\textwidth]{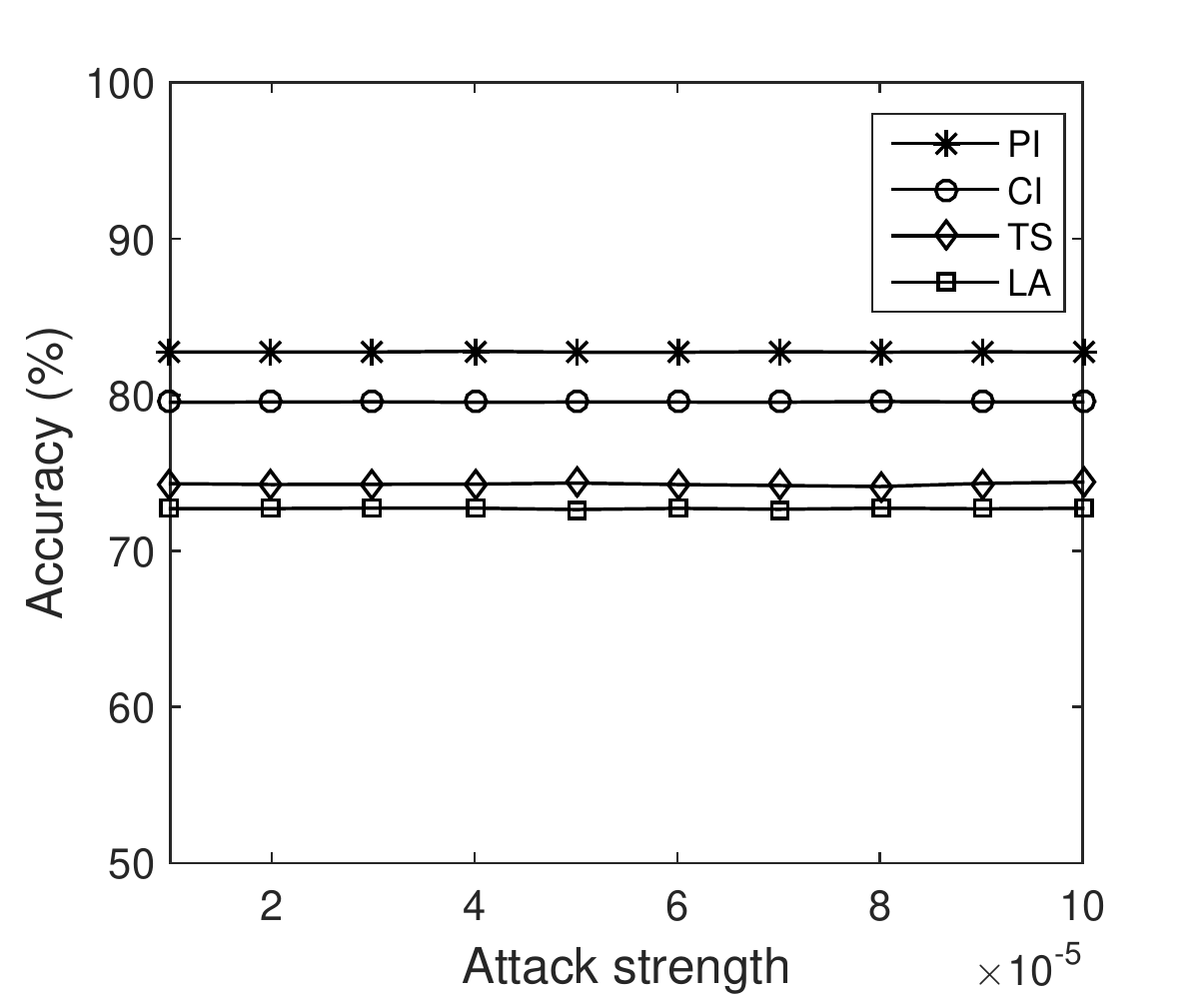}}\hfill
\subfigure[Social-bot influence ranking]{\label{fig:df2-8}\includegraphics[width=0.5\textwidth]{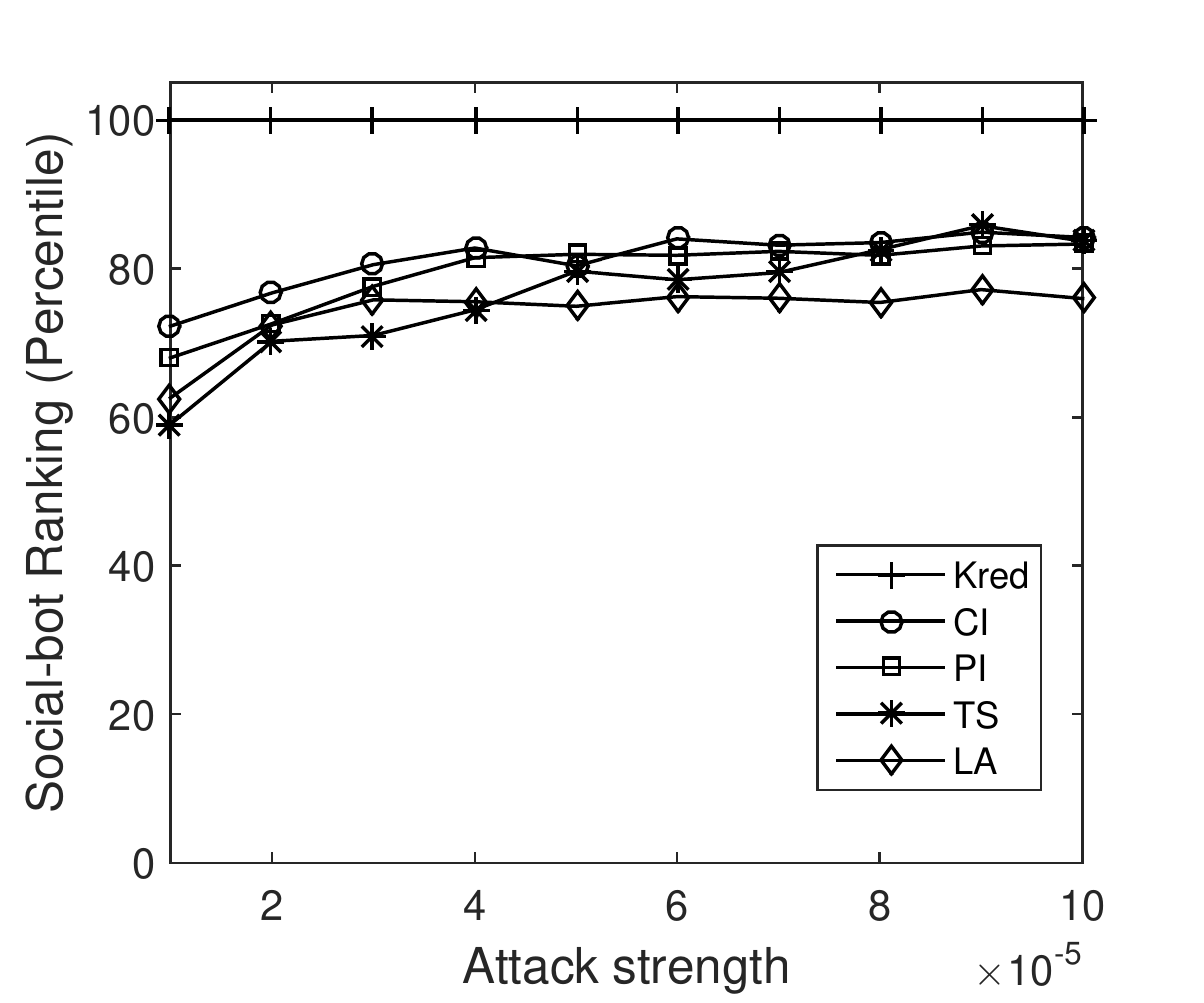}}\hfill
\caption{The performance under the seed-targeting attack.}
\label{fig:df2-attack2}
\end{minipage}
\vspace{-.2in}
\end{figure*}

\subsection{Defense against Digital-influence Manipulation} \label{sec:defend2}
As discussed in Section~\ref{sec:diMani}, all the digital-influence software vendors consider the number of actions and/or the audience size as the major factors in evaluating a user's digital-influence, which makes them vulnerable to the social botnet. To address this vulnerability, we propose a new digital-influence measurement scheme inspired by \cite{GyongCom04}. The key idea is to find sufficient credible users and only use the actions from them to compute digital-influence scores for other users. Specifically, given a social network with user set $V$ and all the actions (including following, retweeting, mentioning, and replying) among the users, we first find a subset $V^*\subseteq V$ of users that are credible.

We then define the digital-influence score of each user $v$ based on the actions from the credible user set $V^*$. Specifically, assume that user $v$ has received $a_j$ actions from each user $j\in V^*$. The digital influence score for $v$ is then given by as
\begin{equation}
\mathsf{score}(v)=\sum_{j\in V^*} f(a_j)
\end{equation}
where
\begin{equation}
 f(a_j)=\begin{cases}a_j & \text{if $a_j=0,1$,}\\
 1 + \lambda \cdot \exp(-\frac{1}{a_j}) &\text{else,}
\end{cases}
\end{equation}
and $\lambda$ is a system parameter that represents the maximum impact of actions from a single user on one's score. In practice, we can set $\lambda=1$ such that a single user could contribute the score by at most 2. It is easy to see that the above score definition takes both the number of actions and the audience size into account, which captures the key ideas behind Klout, Kred, Retweet Rank scores as well our findings in Section~\ref{sec:BotDIM}.

The challenge is then how to find the credible user set $V^*$. To tackle this challenge, we observe that although the actions among the social bots and from the social bots to the legitimate users are unpredictable, legitimate users usually carefully choose whom to interact, so there are much fewer actions from legitimate users to social bots. Based on this observation, we first find a small number of trusted users, which could either be the verified users maintained by Twitter or manually verified.

We then assign each trusted user some initial credits and distribute the credits along the action edges in multiple rounds. In each round, every user with credits keeps one unit of credit for himself and distributes the remaining credits to his neighbors along the action edges. The credit distribution process terminates when no user has extra credits to distribute. Since legitimate users are more likely to receive credits than social bots during the process, every user who has received at least one unit of credit is considered credible.

More specifically, we first construct an action graph for credit distribution. Given the user set $V$ and all their actions during the period $T$, we build a weighted and directed action graph $\mathcal{G}=(V,E)$, where each user corresponds to a vertex in $V$, and there is an arc $e_{ij}$ from user $i$ to user $j$ an edge with the weight $w_{ij}$ if user $i$ has retweeted, replied, or mentioned user $j$ for $w_{ij}$ times during the period $T$. We do not consider the following action because it has been reported that normal users could carelessly follow back whoever has followed them \cite{YangAna12, GhoshUnd12}.

We then design a credit distribution scheme on $\mathcal{G} = (V,E)$, which consists of seeds selection, initial credit assignment, iterative credit distribution, and termination. To initiate credit distribution, we first select a seed set $S$ from $V$ which are trusted users such as the verified users maintained by Twitter or manually verified, and partition the whole graph into a tree of multiple levels, in which the seed users occupy the first level, their one-hop outgoing neighbors occupy the second level, and so on. A node is assigned to the highest level if its incoming neighbors appear at different levels. We then assign each seed $s$ with the amount of credits proportional to the sum of weights of its outgoing edges, i.e., $w_o(s) C_{\textrm{total}}/\sum_{s\in S}w_o(s)$, where $C_{\textrm{total}}$ is the total amount of initial credits, and $w_o(s)$ is the sum of weights of the seed $s$'s all outgoing edges. We then distribute the credit from the seed users along the tree level by level. Specifically, if a user $v$  at the $n$th level has $c(v)$ units of credit, he holds one unit of credit for himself and distributes the remaining $c(v)-1$ units of credit to its outgoing neighbors at the $(n+1)$th level, where the amount of credits received by neighbor $v'$ is proportional to the edge weight $e_{vv'}$. In other words, neighbor $v'$ receives $w_{vv'} (c(v)-1)/ \sum_{u\in \mathcal{O}(v) w_{vu}}$ units of credits. The credits could only be distributed from one level to the immediate higher level and are rounded to the closest integer. The credit distribution terminates if none of the nodes has any extra credit to distribute to its outgoing neighbors.

The choice of $C_{\textrm{total}}$ represents the tradeoff between true and false positive rates.  On the one hand, with a larger $C_{\textrm{total}}$, we could discover more credible users and compute digital-influence scores for more legitimate users at the cost of more social bots obtaining credits and being labelled as credible users. On the other hand, a smaller $C_{\textrm{total}}$ will result in fewer credible users being discovered as well as fewer social bots being labelled as credible users. As in \cite{TranSyb09}, we set $C_{\textrm{total}} = O(\sqrt{|V|})$ to strike a good balance between true and false positive rates. The idea is that at the end of credit distribution, approximately $C_{\textrm{total}}$ users will each hold one unit of credit and be labelled as credible. If we set $C_{\textrm{total}} = O(\sqrt{|V|})$ and the interaction graph is well-connected,  with high probability there exists at least one arc  from some credible users to each of the $|V|$ users while the credit distribution will terminate in $O(\log{|V|})$ levels with decreasing credits per arc such that each attack arc from legitimate users to social bots is expected to receive only $O(1)$ of credits  \cite{TranSyb09}.

\subsubsection{Evaluation}

We follow the approach taken by \cite{YuSyb06, CaoAid12,ZhangTru15} to evaluate the performance of the digital-influence measurement
scheme. Specifically, we first use the Twitter geo-search API \cite{api13} to collect the users in a specific metropolitan area and then crawl the latest 600 tweets of each user to extract the interactions such as retweets, replies, and mentions. We create four datasets from the four representative area in U.S. as shown in Table~\ref{tlb:datasets}.

For each dataset, we build one ground-truth network composed of both legitimate users and social botnets. Specifically, we first build one \emph{legitimate action subgraph} from the interactions among the legitimate users in the that area, where each vertex corresponds with one legitimate user, and an edge from one user to another corresponds to interaction from the former to the latter.
We further create a completely-connected social botnet with the same number of social bots as the legitimate users. We then construct a ground-truth network by connecting the legitimate action subgraph and the social botnet with a number of \emph{attack edges} from legitimate action subgraph to the social botnet. Here we assume that there is no edge from the social botnet to the legitimate action subgraph, which is clearly in favor of the social bots, as the social bots would only lose credits through such edges.

The attack edges are constructed in the following way. Assuming that there are total $g$ edges in the legitimate action subgraph, we create $\omega g$ attack edges with unit weight, where $\omega$ is the ratio representing \emph{attack strength}. As in \cite{ZhangTru15}, we vary the $\omega$ from $10^{-5}$ to $10^{-4}$ of the total edges in each legitimate subgraph. We use ground-truth networks to conduct the credit distribution and find the credible user subset $V^*$.

As in \cite{CaoAid12}, we consider two types of attacks. In the \emph{random attack}, we randomly choose $\omega g$ legitimate users and connect each of them to one randomly chosen social bot with an attack edge. In this case, the attacker is unaware of which users are seed. In the \emph{seed-targeting attack}, we create $\omega g$ attack edges between the social botnet and $\omega g$ users randomly chosen from $100\omega g$ legitimate users with shortest distances to the seeds. This type of attack mimics the case that the attacker tries to acquire positions close to the seed users to gain more credits during credit distribution. For both attacks, we assume that the social botnet can arbitrarily distribute received credits internally. For example, suppose that the social botnet receives $C_{\textrm{bots}}$ credits after the credit distribution, they distribute these credits to $C_{\textrm{bots}}$ social bots each with one credit, such that all $C_{\textrm{bots}}$ bots become credible users.


We evaluate the performance of the proposed digital-influence measurement scheme using the following metrics. 
\begin{itemize}
\item \emph{Top-$K$-percent accuracy}. Let $\mathcal{U}_1$ be the list of legitimate users ranked by their total numbers of incoming interactions. Also let $\mathcal{U}_2$ be the list of legitimate users ranked by their digital influence scores obtained by applying the proposed digital-influence measurement scheme to the ground-truth network. Further let $\mathcal{U}_1(K)$ and $\mathcal{U}_2(K)$  be the top-$K$-percent users in $\mathcal{U}_1$ and $\mathcal{U}_2$, respectively. The top-$K$-percent accuracy is defined as $\frac{|\mathcal{U}_1(K) \cap \mathcal{U}_2(K)|}{|\mathcal{U}_1(K)|}$. The more accurate of the proposed scheme, the higher the ratio of common users in these two top-K lists, and vice versa.
\item \emph{Social-bot influence ranking}. Let bot $b$ be the bot with the highest digital influence score output by the proposed digital-influence measurement scheme. Social-bot influence ranking is defined as $b$'s rank in percentile in $\mathcal{U}_1$. The lower the bot's percentile, the higher resilience to the social botnet, and vice versa.
\end{itemize}

Fig.~\ref{fig:df2-attack1} and Fig.~\ref{fig:df2-attack2} show the \emph{top-$10$-percent accuracy} and the \emph{social-bot influence ranking} under two types of attacks with attack strength varying from $10^{-5}$ to $10^{-4}$. In general, we see similar trend in all scenarios for all the four datasets, meaning that the results are consistent across different datasets. Moreover, in all cases, the accuracy is always larger than 70\%, meaning that the proposed digital-influence measurement scheme can discover 70\% of top-10\% influential users under the social botnet attack. We can also see in Fig.~\ref{fig:df2-3} and Fig.~\ref{fig:df2-7} that the the accuracy is insensitive to the change of attack strength. The reason is that the top influential users attract actions from many legitimate users, who have higher chances to acquire the credits and become credible users in the presence of attack edges. In contrast, as shown in in Fig.~\ref{fig:df2-4} and Fig.~\ref{fig:df2-8}, the social-bot influence ranking increases as the attack strength increases under both types of attacks. This is expected, because as attack strength increases, more social bots will acquire credits to become credible users and affect target bots' influence scores. We can also see from Figs.~\ref{fig:df2-4} and \ref{fig:df2-8} that the social-bot influence ranking is below top 40\% and top 20\% under the random attack and seed-targeting attack, respectively.

Fig.~\ref{fig:df2-4} and Fig.~\ref{fig:df2-8} also compare the proposed defense with the existing influence measurement vendor Kred.\footnote{we choose Kred because only Kred has publish its influence score model on \url{http://kred.com/rules}.} As we can see, the bot's influence score could always rank at the first position because we assumed that there are infinite actions between any two bots in the botnet. In contrast, the proposed scheme could effectively defend against the manipulation from social botnets.


\begin{figure}[t]
\centering
    \subfigure[Random Attack]{\label{fig:df2-topk-3}\includegraphics[width=0.24\textwidth]{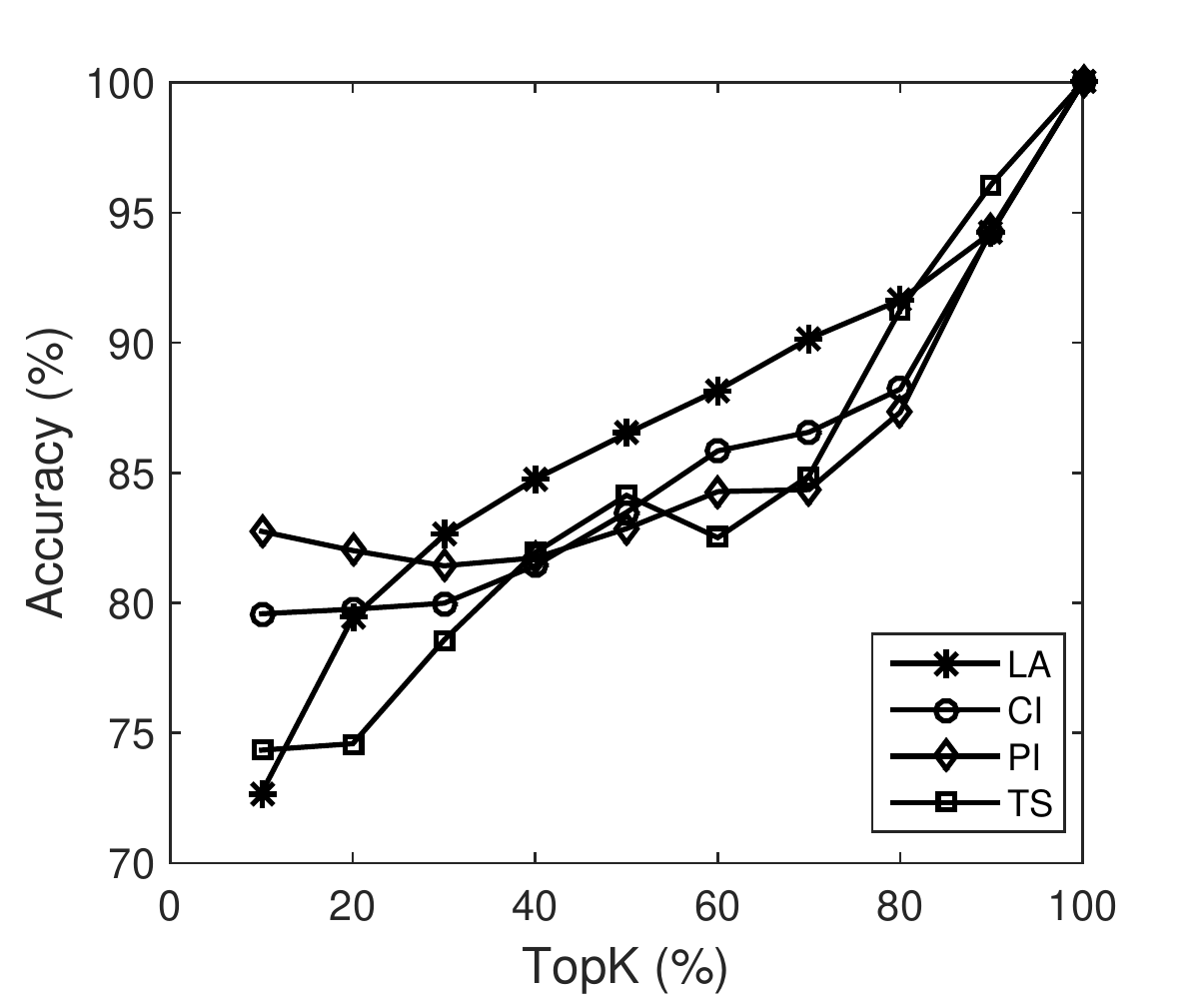}} \hfill
    \subfigure[Seed-targeting Attack]{\label{fig:df2-topk-4}\includegraphics[width=0.24\textwidth]{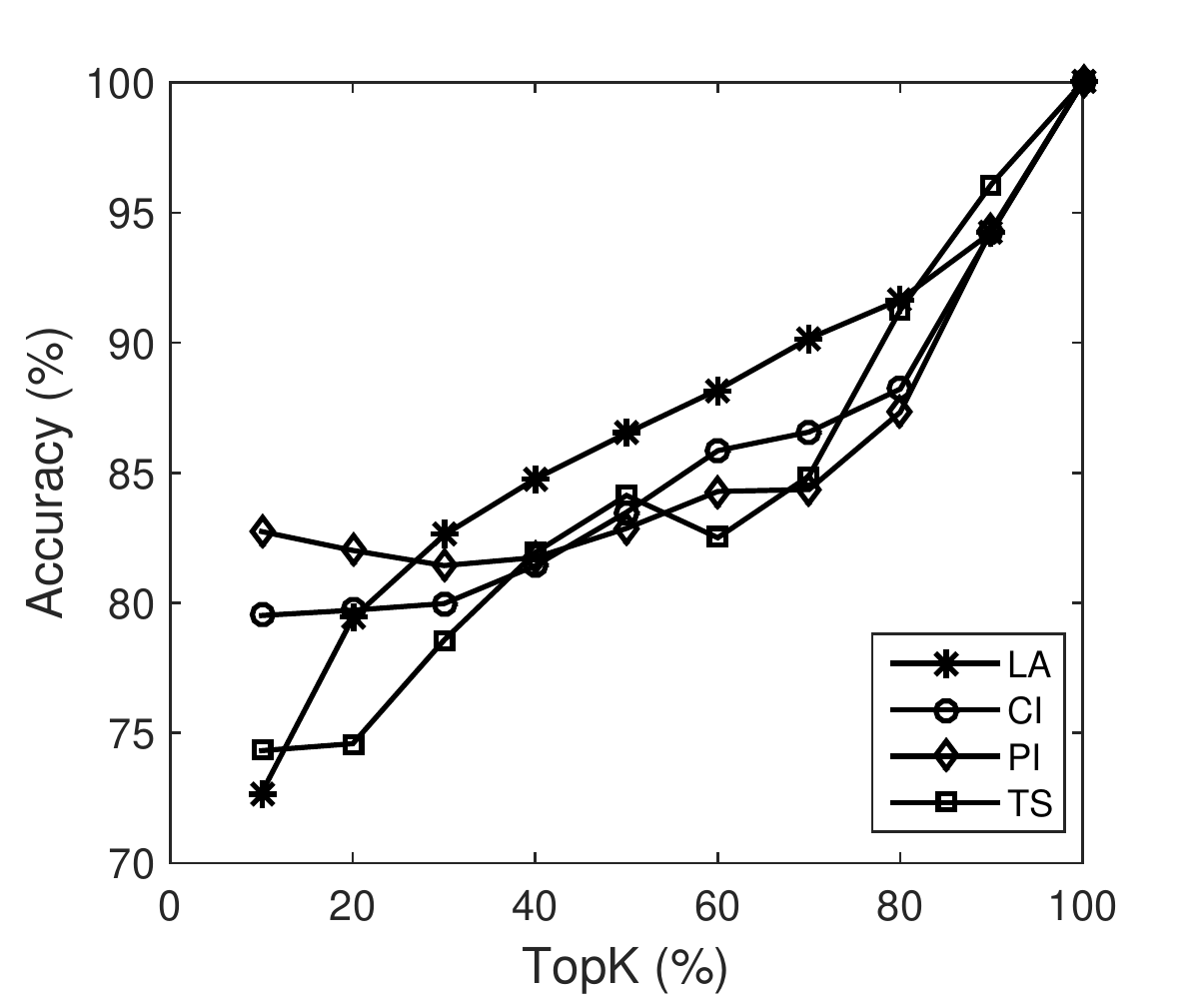}} \hfill
    \caption{The impact of $K$ on the top-$K$-percent accuracy.}
\label{fig:df2-topk}
\vspace{-.2in}
\end{figure}


Fig.~\ref{fig:df2-topk} shows the impact of $K$ on the top-$K$-percent accuracy, which generally increases until to 100\% when $K$ increases from 10 to 100. Overall, the accuracy is higher than 70\% in the figures, and $K=10$ in previous experiments is representative among the worst cases.

In summary, experiment results show that the proposed digital-influence measurement scheme is resilient to the social botnet attack.

\section{Related Work}\label{sec:relatedWork}

In this section, we discuss the prior work tightly related to our work in this paper.

The social botnet has received attention only recently. Ferrara \emph{et al.} summarized the research on social bots in \cite{FerraRis14}.  Boshmaf \emph{et al.} showed that a social botnet is very effective in connecting to many random or targeted Facebook users (i.e., large-scale infiltration)  \cite{BoshmSoc11}. The work in \cite{YangTas14, FreitRev14} shows how the spammers become smarter to embed themselves into Twitter. Messias \emph{et al.} used two sicalbots to manipulate the Klout and Twitalyzer scores by following and tweeting actions\cite{MessiYou13}. Our preliminary work \cite{ZhangOn13} was independently done from \cite{MessiYou13} and shows how social bots could cooperate to manipulate the influence scores of Klout, Kred, and Retweet Rank.


There is a rich literature on spam detection in OSNs. The first line of work such as \cite{ZhangDet11,SongSpa11,ThomaDes11, GaoTow12, StrinDet10, LeeUnc10, YangDie11, BenevDet10, WangDon10, LeeWar12} considers independent spam bots and comes up with different methods to characterize and detect them. For example, Song \emph{et al.} found that the distance between the sender and receiver of a message is larger for spammers than legitimate users \cite{SongSpa11}. Zhang \emph{et al.} found that automated spammers usually show some temporal pattern which can be used for detection \cite{ZhangDet11}. Another features and the corresponding machine learning effort can be found in \cite{ThomaDes11, GaoTow12, StrinDet10, LeeUnc10, YangDie11, BenevDet10, WangDon10, LeeWar12}. Some of these results such as \cite{SongSpa11} have been incorporated into our design constraints in \S~\ref{sec:designConstraint}.

The second line of work such as \cite{GaoDet10,GrierSpa10,ThomaSus11, Chudet12,YangAna12,GhoshUnd12} focuses on characterizing and detecting organized spam campaigns launched by an army of spam bots. We discover a new method for effective spam campaign in this paper, and whether the results in \cite{GaoDet10,GrierSpa10,ThomaSus11,Chudet12,YangAna12,GhoshUnd12} can be directly or adapted to detect our method is certainly challenging and worthy of further investigation. Moreover, spam bots are evolving towards more intelligence. Yang \emph{et al.} found that instead of separate communities, spam bots embedded themselves into the network by building small-world-like networks between them and also maintaining tight links with external legitimate users\cite{YangAna12}. Meanwhile, Ghosh \emph{et al.} discovered the similarly rich connections between spam bots and the legitimate users\cite{GhoshUnd12}. Our work is consistent with this trending and explore the new attacks by using the spam bots' growing intelligence.

There are effective solutions such as \cite{CaoAid12, ViswaExp12, ViswaAna10, TranSyb09, YuSyb10, YuSyb06} to detecting Sybil accounts in distributed systems under the control of a single attacker. These solutions commonly assume that the link between two accounts corresponds to a real social relationship difficult to establish and impossible to forge. Moreover, all of these system assume that the connection is undirected. In this paper, we designed the defense scheme for directed and weighted Twitter network by using the similar observation that the amount of interactions from a legitimate user to a social bot is usually far less than that in the reverse direction in \S~\ref{sec:defend2}. We also found that using the interaction network yields better defense performance than the following networks. Although \S~\ref{sec:defend2} has illustrate an example by using the trustworthiness of interactions to defend against the specific digital-influence manipulation attack, we can use the same observation but different methods to identify the social botnets and then eliminate them from the microblogging system.

Also related is the research on computing the digital influence score \cite{ChaMea10, WengTwi10, BakshEve11}. For example, Cha \emph{et al.}, compared three influence ranking lists in Twitter by counting the number of retweets, mentions and followers and found that the influenced defined by the number of followers is very different with the other two metrics \cite{ChaMea10}. Bakshy \emph{et al.} \cite{BakshEve11} proposed to measure user influence based on his ability to post the tweets that generates a cascade of retweets. This line of research does not consider the impact of the social botnet.

\section{Conclusion}\label{sec:conclusion}

In this paper, we firmly validated the efficacy and merits of botnet-based spam distribution and digital-influence manipulation on Twitter through thorough experiments and trace-driven simulations. We also propose the countermeasures corresponding to these two attacks and demonstrate their effectiveness.

\bibliographystyle{IEEETran}

\end{document}